\documentclass[reprint,showpacs,aps,pra,longbibliography]{revtex4-1}
\usepackage[latin9]{inputenc}
\usepackage{bm}
\usepackage{amsmath}
\usepackage{amssymb}
\usepackage{graphics}
\usepackage{bm}
\usepackage{bbm}
\usepackage{graphicx}
\usepackage{amsthm}
\makeatother

\begin{document}

\title{Conserved operators and exact conditions for pair  condensation}

\author{Federico Petrovich$^1$ and R. Rossignoli$^{1,2}$}
\affiliation{$^{1}$Instituto de F\'{\i}sica  La Plata, CONICET, and Depto.\  de F\'{\i}sica, Facultad de Ciencias Exactas, Universidad Nacional de La Plata, C.C. 67, La Plata (1900), Argentina\\
 $^{2}$Comisi\'on de Investigaciones Cient\'{\i}ficas (CIC), La Plata (1900), Argentina}
\begin{abstract}
We determine the necessary and sufficient conditions which ensure 
that an $N=2m$-particle fermionic  or bosonic state  has the  form $|\Psi\rangle\propto(A^{\dagger})^{m}|0\rangle$,
where  $A^{\dagger}=\tfrac{1}{2}\sum_{i,j}A_{ij}c_{i}^{\dagger}c_{j}^{\dagger}$
is a general pair creation operator. These conditions can be cast as an eigenvalue equation for a modified two-body density matrix, and enable an exact reconstruction of the  operator $A^\dag$, providing  as well  a measure of the proximity of a given state to an exact pair condensate.  Through a covariance-based formalism, it is also shown that such states are fully characterized by a set of $L$ ``conserved'' one-body operators which have $|\Psi\rangle$ as exact eigenstate, with $L$ determined just by the single particle space dimension involved. The whole set of two-body Hamiltonians having $|\Psi\rangle$  
as exact eigenstate is in this way determined, while a general subset having $|\Psi\rangle$ as  nondegenerate ground state is also  identified.  Extension to states $\propto f(A^\dag)|0\rangle$ with $f$ an arbitrary function is also discussed. 
\end{abstract}
\maketitle

\section{Introduction\label{I}}
The exact eigenstates of interacting many-body  Hamiltonians have normally a complex entangled structure \cite{AF.08}.  
Approximate descriptions  based on special simple forms of the many-body state have therefore been introduced from the very beginning of quantum mechanics,  starting from mean field (MF)-type  approaches  based on independent particle or quasiparticle states  like  Slater-determinants (SD) or BCS-type states for fermions \cite{H.28,F.30,bcs.57,RS.80}. 
More complex  approaches based on  projected (i.e.\ symmetry-restored) MF states, when the latter break some relevant symmetry of the Hamiltonian \cite{RS.80}, as well as bosonic-like ans\"atze based on particle pairs, such as the general random phase approximation (RPA) scheme \cite{R.68,RS.80}, were also introduced in early stages, followed more recently by other  schemes \cite{Scw.05,VMC.08}.  

In particular, the so-called pair  condensates \cite{LLJ.22},   also denoted as coboson condensates \cite{CKL.05,CRBMC.19} (or previously as antisymmetrized geminal powers \cite{AJC.65}),   provide an adequate  approach for describing some relevant even $N=2m$-particle states  in different contexts \cite{LLJ.22,CKL.05,CRBMC.19,AJC.65,RSC.91,CBD.08,CB.10,CSC.16,CB.02,Comb.162,COW.10,JCMC.23}. These states have the general form $|\Psi\rangle\propto (A^\dag)^m|0\rangle$, with $A^\dag$ a general pair creation operator, normally generating  a ``collective''  entangled pair state when applied on the vacuum. Thus,   $|\Psi\rangle$ can be considered as a condensate of $m$ pairs, which behave approximately as bosons  due to the ensuing integer spin of the pair. These states also emerge naturally as particle number projected quasiparticle vacua, as the latter can  be expressed as 
$\propto e^{-\alpha A^\dag}|0\rangle$ for both fermions or bosons  \cite{RS.80}, when having positive number parity, hence yielding a $2m$-particle component $\propto (A^\dag)^m|0\rangle$.  For instance, a particle number projected BCS  or Hartree-Fock-Bogoliubov fermionic state is of the previous form \cite{RS.80}. 
Hence, they arise in systems with pairing interactions, where they can become exact eigenstates in certain limits  or at certain special points, as will be discussed. 

The first goal of this work is to characterize these states through a novel scheme based on ``conserved'' operators, i.e.\ operators  which have these states as exact eigenstates. Accordingly, we start from a general quantum covariance-based approach,  which allows one to identify the set of conserved operators of a certain class, like e.g.\ one-body operators,  
 inspired by a recent treatment of eigenstate separability for  systems of distinguishable components \cite{PRC.24,PCR.22}. We will then  show 
that general pair condensates  $|\Psi\rangle$, which can be regarded  as ``uniformly separable'' at the pair level (in the sense of being a power of a single pair creation operator applied to the vacuum), 
are fully characterized by a fixed number of exactly conserved one-body operators, which depend just on the single particle (sp) space dimension involved and not on the number of pairs. This number is in fact the highest among states covering the full sp space (without fully occupied levels in the fermion case), reflecting their special structure. 
From this set the most general two-body Hamiltonian having the pair  condensate as eigenstate will also be obtained, together with a general class of Hamiltonians which  have it as nondegenerate ground state (GS). 

From the previous formalism, we are then able to determine an  exact necessary and sufficient condition  which ensures
that a given state $|\Psi\rangle$ of $N=2m$ fermions or bosons is an exact  pair condensate, which is our second aim. This condition involves just an eigenvalue equation for a modified two-body density matrix  (DM), and  yields the corresponding exact pair creation operator $A^\dag$ determining the state, thus enabling  its  exact reconstruction. In addition, it also provides a simple measure of the proximity of a given state to a pair condensate, together with a ``best'' pair condensate approximation. 
Our treatment is exact and hence does not rely on any bosonic assumption or approximation to the state, yielding  a unified characterization of both fermionic or bosonic pair condensates.  
 The extension to pure or mixed states with no fixed particle number, and to neighboring odd states,  is also provided. 
 The formalism and  main results are  discussed in section \ref{II}, while illustrative examples  are provided  in  section \ref{III}. Appendices contain proofs and additional details. Conclusions are finally drawn in \ref{IV}.  

\section{Formalism\label{II}}

\subsection{State of the problem}

We start from a set of $n$ fermion or boson creation and annihilation
operators $c_i^\dagger,c_i$ satisfying $[c_i,c_j^\dagger]_{\pm}=\delta_{ij}$ and $[c_i,c_j]_{\pm}=0=[c_i^\dagger,c_j^\dagger]_{\pm}$, where the upper
sign will always correspond to fermions and the lower one to bosons,
with $[a,b]_{\pm}=ab\pm ba$. We want  to  determine the necessary
and sufficient conditions for which an $N=2m$-particle state has
the form
\begin{equation}
|\Psi\rangle=|m\rangle_2:=\tfrac{1}{\sqrt{\mathcal{N}_m}}(A^{\dagger})^{m}|0\rangle\,,
\label{State}
\end{equation}
where 
\begin{equation}
A^{\dagger}=\tfrac{1}{2}\sum_{i,j}A_{ij}c_{i}^{\dagger}c_{j}^{\dagger}\,,\label{A}
\end{equation}
is a general pair creation operator, with $A_{ij}=\mp A_{ji}$, $\langle0|AA^{\dagger}|0\rangle=\frac{1}{2}{\rm Tr}[{\bf A}^\dag {\bf A}]= 1$ (${\bf A}$ is the  matrix of elements $A_{ij}$) and $\mathcal{N}_m=\langle0|A^{m}A^{\dagger m}|0\rangle$. We can always
write $A^{\dagger}$ in the Schmidt-like diagonal form \cite{ESBL.02} $\!\!$
\begin{subequations}
\begin{eqnarray}
A^{\dagger}&=&\sum_{k=1}^{n/2}\sigma_{k}a_{k}^{\dagger}a_{\bar{k}}^{\dagger},
\label{Schmidtfer} \\
A^{\dagger}&=&\tfrac{1}{\sqrt{2}} \sum_{k=1}^{n}\sigma_{k}b_{k}^{\dagger2}, \label{Schmidtbos}
\end{eqnarray}
\label{Schmidt}
\end{subequations}
$\!\!$where \eqref{Schmidtfer} corresponds to fermions (here we can assume
$n$ even), \eqref{Schmidtbos} to bosons, with $\sum_{k}|\sigma_k|^2=1$ in both cases.  

Without loss of generality, we can assume $\sigma_{k}\neq 0\ \forall\ k$, by setting $n$ as the rank of ${\bf A}$, i.e.\ as the dimension of the sp space  occupied by the condensate \eqref{State}, such that ${\bf A}$ is nonsingular. We can  also assume  $\sigma_k\in\mathbb{R}_+$ $\forall\,k$  by adjusting the phase of the $a^\dag_k$ or  $b^\dag_k$, in which case $\sigma_k$ ($\sqrt{2}\sigma_k$)  are the singular values of ${\bf A}$. The operators $a^\dag_k$ ($b^\dag_k$) are unitarily related to the $c_i^\dag$ \cite{ESBL.02}, and in the fermionic case (${\bf A}$ antisymmetric) the singular values are always twofold-degenerate, with the diagonalizing transformation defining a set of orthogonal sp states ($k,\bar k$) for $k=1,\ldots,n/2$. 

For fermions we also have $0\leq m\leq n/2$ (as $(A^\dag)^m = 0$ for $m>n/2$), with $|m = \frac{n}{2}\rangle_2 = |\bar{0}\rangle=(\prod_{k=1}^{n/2}a^\dag_k a^\dag_{\bar k})|0\rangle$ the ``fully occupied'' state, i.e.\ that having all $n$ sp states   with maximum occupancy $1$ (a SD),   $\forall\,{\bf A}$ of rank $n$. 
 
If $\sigma_k = \tfrac{1}{\sqrt{n/2}}$ $(\frac{1}{\sqrt{n}})$ $\forall\,k$ for fermions (bosons), both Eqs.\ \eqref{Schmidt} lead to a perfect ladder operator $A_0^\dag$ satisfying $[A_{0},A_{0}^{\dagger}]=1\mp 2\hat{N}/n$ with $\hat{N}=\sum_{i}c_{i}^{\dagger}c_{i}$ the number operator, which has special properties (see App.\ \ref{ApA}). In the general case, this relation is generalized to
\begin{equation}
[\bar{A},A^\dag] = 1\mp 2\hat{N}/n,
\label{comAbar}
\end{equation}
where $\bar A$ is the ``dual'' pair annihilation operator 
\begin{equation}
\bar{A}=\tfrac{1}{n}\sum_{i,j}A^{-1}_{ij}c_i c_j\,,
\label{Abar}
\end{equation}
 which coincides with $A=(A^\dag)^\dag$  in the uniform case only. 
 
As a first related result, we prove in App.\ \ref{ApA} the following Proposition for fermions: 

{\bf Proposition 1.} {\it The state \eqref{State} can be also written as 
\begin{equation}
|m\rangle_2=\tfrac{1}{\sqrt{\mathcal{\bar{N}}_m}}(\bar{A})^{\frac{n}{2}-m}|\bar{0}\rangle,
\label{Statebis}
\end{equation}
where $|\bar{0}\rangle$ is the previously defined fully occupied state and $\bar{A}$ the operator \eqref{Abar}, such that any $N=2m$-particle fermionic pair condensate in an $n$-dimensional sp space can be also cast as an $\bar{N}=\frac{n}{2}-m$-hole pair condensate} with respect to $|\bar 0\rangle$. 

Then, since any $N=2$-particle fermionic state obviously has the form \eqref{State} for $m=1$, we can claim that any $n-2$-particle
fermionic state $|\Psi\rangle$ can also be written in the form \eqref{State} for $m=n/2-1$, as it is a two-hole state with respect to $|\bar{0}\rangle$. For $N\geq 4$, with  $n\geq  N+4$ ($n\geq 2$) in the fermionic (bosonic) case, 
a general state is obviously not necessarily of the form \eqref{State}. 

We will also consider the conditions for more general pure states of the form 
\begin{equation}
|\Psi_A \rangle = f(A^\dag) |0\rangle=\sum_{m}\alpha_m|m\rangle_2\,,
\label{State2}
\end{equation}
where $f(x)=\sum_m \alpha_m x^m$  is an arbitrary function, and also the mixed states  
\begin{equation}
\rho_A=\sum_m \alpha_{mm'}|m\rangle_2\langle m'|\,,
\label{State2m}
\end{equation}
which include in particular the pure case \eqref{State2} ($\alpha_{mm'}=\alpha_m\alpha_{m'}^*$) and the diagonal  case $\alpha_{mm'}=p_m\delta_{mm'}$, i.e.,  
$\rho_A^d=\sum_m p_m|m\rangle_2\langle m|$, which will be always assumed normalized  
Finally, we will  discuss the conditions for  neighboring  odd-number states $|\Psi_{\rm odd} \rangle \propto c^\dag_i|m\rangle_2$ and $c_i|m\rangle_2$ for arbitrary $c_i^\dag$, $c_i$. 

\subsection{Conserved quantities and covariance matrix}

Our approach is based on first identifying this family of states through the set of  ``conserved'' operators $Q_{\alpha}$ of a certain class, satisfying \begin{equation}
Q_{\alpha}|m\rangle_2=\lambda_{\alpha}|m\rangle_2\,,\label{QC}
\end{equation}
such that $\langle Q_\alpha^\dag Q_\alpha\rangle-\langle Q_\alpha^\dag\rangle\langle  Q_\alpha\rangle=0$  
for  $\langle O\rangle={_2\langle}m|O|m\rangle_2$. These operators can   then be obtained from the nullspace of the pertinent covariance matrix ${\bf C}$,  of elements 
\begin{equation}
C_{\mu\nu}=\langle O_{\mu}^{\dagger}O_{\nu}\rangle-\langle O_{\mu}^{\dagger}\rangle\langle O_{\nu}\rangle\,,
\end{equation}
for $O_{\mu}, O_\nu$ belonging to a certain set $\mathcal{B}$.   Its nullspace is composed of vectors $\bm{h}_{\alpha}$, $1\leq\alpha\leq L$, such that ${\bf C}\bm h_\alpha=\bm 0$, implying   $\langle Q_\alpha^\dag Q_\alpha\rangle-\langle Q_\alpha^\dag\rangle\langle Q_\alpha\rangle=\bm h_\alpha^\dag\,{\bf C}\bm h_\alpha=0$ for $Q_\alpha=\sum_\mu h_\alpha^\mu O_\mu$. For averages with respect to a pure state $|\psi\rangle$, always assumed normalized, this implies 
$Q_\alpha|\psi\rangle=\lambda_\alpha|\psi\rangle$ \cite{PRC.24}. Thus, the subspace of conserved operators $Q_\alpha\in{\cal B}$ associated to a  state $|\psi\rangle$ is fully determined by the nullspace of ${\bf C}$. Notice that if $Q_{\alpha}$ and $Q_{\alpha'}$ are both conserved, so will be $[Q_\alpha,Q_{\alpha'}]$, implying that the full set of conserved operators is always closed under commutation. 

For systems of  indistinguishable particles,  the $Q_{\alpha}$ are  polynomials in $c_i, c_i^\dag$, and the set ${\cal B}$ may refer e.g.\ to one-body operators, or pair creation operators, etc. While the latter commute among themselves,  the former are closed under commutation, such that the set of conserved one-body operators associated to a given state $|\psi\rangle$ form a closed subalgebra of the full set. It also defines a set of one-body transformations $U_Q=\exp[\sum_\alpha \gamma^\alpha Q_\alpha]$, not necessarily unitary, which leave $|\psi\rangle$ invariant except for a constant: 
$U_Q|\psi\rangle=e^{\sum_\alpha \gamma^\alpha \lambda_\alpha}|\psi\rangle$ 
if $Q_\alpha|\psi\rangle=\lambda_\alpha|\psi\rangle$ $\forall\, Q_\alpha$. 

If $|\psi \rangle$ has definite particle number, $\lambda_\alpha=0$ for all  $Q_\alpha$ satisfying \eqref{QC}  which do not conserve the number of particles ($[Q_\alpha,\hat{N}]\neq 0$). Moreover, from a given set of  conserved operators $Q_\alpha$ of a certain class, not necessarily hermitian, we may always construct the hermitian conserved quadratic ``Hamiltonian'' 
\begin{eqnarray}
H_Q=\tfrac{1}{2}\sum_{\alpha,\beta}V_{\alpha\beta}\tilde{Q}_{\alpha}^{\dagger}\tilde{Q}_{\beta},
\label{Hamiltonian}
\end{eqnarray}
where $\tilde{Q}_{\alpha}:= Q_\alpha - \langle Q_\alpha\rangle=Q_\alpha-\lambda_\alpha$ satisfies  $\tilde Q_\alpha|\psi\rangle=0$ and ${\bf V}={\bf V}^{\dagger}$ (${\bf V}$ is the matrix of coefficients
$V_{\alpha\beta}$). Hence  $H_Q$ will 
 have  $|\psi\rangle$ as eigenstate with zero energy: $H_Q|\psi\rangle=0$. Moreover, if ${\bf V}$ is positive definite, $H_Q$ is positive semidefinite (as diagonlization of ${\bf V}$ leads to $H_Q=\sum_\nu\Lambda_\nu \tilde O^\dag_\nu \tilde O_\nu$ with $\Lambda_\nu>0$ the eigenvalues of ${\bf V}$ and $\tilde O^\dag_\nu \tilde O_\nu$ positive semidefinite operators), implying $\langle H_Q\rangle\geq 0$  and hence $|\psi\rangle$  {\it a GS} of $H_Q$ as $\langle\psi|H_Q|\psi\rangle=0$.  If the $Q_{\alpha}$ define the state univocally, $|\psi\rangle$ will be a {\it non-degenerate} GS of $H_Q$. 

We can  also construct the more general conserved operator (not necessarily hermitian)   
\begin{equation}
H'_Q=\sum_\alpha h_\alpha Q_\alpha+\sum_{\mu,\alpha}V_{\mu\alpha} O_\mu \tilde Q_\alpha\,,\label{Hamiltonianp} 
\end{equation}
where $O_\mu$ are arbitrary operators and $h_\alpha, V_{\mu\alpha}$ arbitrary parameters. It satisfies $H'_Q|\psi\rangle=(\sum_\alpha h_\alpha \lambda_\alpha)|\psi\rangle$ irrespective of the $V_{\mu\alpha}$,  
since $|\psi\rangle$ behaves as    a ``vacuum'' for all centered conserved 
operators $\tilde Q_\alpha$. 

For example, a standard boson condensate 
\begin{equation}
|m\rangle_1 = \tfrac{1}{\sqrt{m!}} (b^\dag)^m |0\rangle\,,
\label{Statem1}
\end{equation}
where $b^\dag= \sum_i \alpha_i c_i^\dag$ is an arbitrary single boson creation operator ($\sum_i |\alpha_i|^2=1$) and $m\geq 1$,   can be  recognized through the covariance matrix of the operators $c_i$, 
\begin{equation}
C^{(1,0)}_{ij} = \langle c_i^\dag c_j \rangle=\rho^{(1)}_{ji}\,,
\label{C10}
\end{equation}
(for states with definite particle number), which is just the transpose of the one-body DM ${\bm \rho}^{(1)}$. It has clearly rank 1 in the state \eqref{Statem1} ($_1\langle m|b^\dag_k b_l|m\rangle_1=m\delta_{kl}\delta_{k1}$ for the natural operators $b^\dag_k=\sum_i
\alpha_{ki}c^\dag_i$ satisfying $[b_k,b^\dag_{k'}]=\delta_{kk'}$ with $b_1^\dag=b^\dag$). And for states with definite particle number, ${\bm \rho}^{(1)}$ has rank $1$ iff the state has the form \eqref{Statem1}. 

Accordingly, these states can be fully characterized by the $n-1$ conserved operators $b_k$, $k=2,\ldots,n$, satisfying $b_k|m\rangle_1=0$, associated to the nullspace of $\bm\rho^{(1)}$. The ensuing conserved Hamiltonian \eqref{Hamiltonian} becomes the one-body operator $H=\sum_{k,l\geq 2} V_{kl} b_k^\dag b_l$, which for $V_{kl} = \delta_{kl}$ is just \begin{equation}H_b=\sum_{k=2}^{n} b_k^\dag b_k = \hat{N} - \hat N_b\,,\label{Hb}\end{equation}
where $\hat N_b=b^\dag b$. 

On the other hand, for a  typical random state (with definite particle number $N\geq 2$ \footnote{We mean a normalized $N$-particle state chosen randomly in the pertinent Hilbert space (assumed of finite dimension) according to the uniform Haar measure \cite{Mel.24}.}) 
there is normally no conserved operator linear in the $c_i$, i.e.\  $\bm \rho^{(1)}$ (or ${\bf C}^{(1,0)}$) has full rank, as all sp states  have  nonzero average occupation in any sp basis.  
Besides, for bosons there are never conserved operators linear in the $c_i^\dag$ either, since a  boson creation operator has no eigenvector. This property can be here easily verified 
as $b b^\dag=1+b^\dag b$ is obviously positive definite for any $b$ linear in the operators $c_i$,  implying that the  covariance matrix ${\bf C}^{(0,1)}$,  
of elements 
$C_{ij}^{(0,1)}=\langle c_i c^\dag_j\rangle$ for states with definite $N$,  
is always positive definite  (the same holds for general states with no fixed boson number, replacing $c_i\rightarrow c_i-\langle c_i\rangle$).

\subsection{Conserved quantities of  pair condensates\label{IIC}}

For the state \eqref{State}, with $m\geq 1$ for bosons  and $1\leq m\leq n/2-1$, $n\geq 4$ for fermions, the covariance matrix \eqref{C10} (and hence $\bm\rho^{(1)}$) is diagonal in the natural sp basis determined by $a_k^\dag$, $a^\dag_{\bar k}$  ($b_k^\dag$ in the boson case), and positive definite if all $\sigma_k$ are  non-zero, since all sp levels are occupied:
$\langle a^\dag_k a_l\rangle=\langle a^\dag_{\bar k} a_{\bar l}\rangle=\delta_{kl}f_k$  for fermions, with $\langle a^\dag_k a_{\bar l}\rangle=0$, while $\langle b^\dag_k b_l\rangle=f_k\delta_{kl}$ for bosons, with $f_k>0$ (and $f_k<1$ for fermions) $\forall\,k$. Hence, we cannot use it for recognizing this state, as many other states can share the same $\bm\rho^{(1)}$ \cite{CR.24}. 

Then, it is expected that the states of the form
\eqref{State} can be identified through conserved quantities
bilinear in $c_{i}$ and $c_{i}^{\dagger}$, i.e.\ one-body operators, or eventually quadratic in $c_i$ or $c_i^\dag$. The covariance matrices for  these three kinds of operators are,  assuming definite particle number, 
\begin{subequations}
\begin{eqnarray}
C_{ij,i'j'}^{(1,1)} &=& \langle c_j^\dag c_i c_{i'}^\dag c_{j'} \rangle -  \langle c_j^\dag c_i \rangle  \langle c_{i'}^\dag c_{j'} \rangle, \label{C11} \\
C_{ij,i'j'}^{(2,0)} &=& \langle c_{i}^{\dagger}c_{j}^{\dagger}c_{j'}c_{i'}\rangle = \rho^{(2)}_{i'j',ij}, \label{C20}\\
C_{ij,i'j'}^{(0,2)} &=& \langle c_{j}c_{i}c_{i'}^{\dagger}c_{j'}^{\dagger}\rangle = \bar{\rho}^{(2)}_{ij,i'j'}\,,\label{C02}
\end{eqnarray}
\label{16}
\end{subequations}
$\!\!$where $\bm\rho^{(2)}$ is the two-body DM \cite{CR.24,Yang.62}. All averages in Eqs.\ \eqref{16} can be obtained from $\bm\rho^{(1)}$ and $\bm\rho^{(2)}$. 

We start with the matrix \eqref{C11}. In App.\ \ref{ApB} we prove the following result:\\

{\bf Theorem 1}. {\it For any $m\geq 1$, with $m\leq n/2-1$ for fermions, the  covariance matrix \eqref{C11} in the state \eqref{State} is singular, having  a nullspace of dimension 
\begin{equation}
L_n=\tfrac{n(n \pm 1)}{2}+1\,,
\label{L}
\end{equation} 
implying  $L_n$ linearly independent conserved one-body operators, given by the number operator $\hat N$, $\hat N|m\rangle_2=2m|m\rangle_2$, and the $L_n-1$ operators 
\begin{equation}
Q_{ij}=(\bm{c}^{\dagger}{\bf A}^t)_i c_j \pm(\bm{c}^{\dagger}{\bf A}^t)_{j} c_i \,,
\label{Qconserved}
\end{equation}
for $i\leq j$ ($i<j$)  for fermions (bosons), satisfying 
\begin{equation}
Q_{ij}|m\rangle_2=0\,.\label{19}
\end{equation} 
They define the state univocally, such that 
$\{Q_{ij}|\Psi\rangle = 0\, \forall\,i,j,  
\hat{N} |\Psi\rangle = 2m |\Psi \rangle\}$ iff $|\Psi \rangle$ has the form \eqref{State}.}

Explicitly, $Q_{ij}=\sum_l c^\dag_l(A_{il}c_j\pm A_{jl}c_i)$, forming a  closed set under commutation, as shown in App.\ \ref{ApB} (Eq.\ \eqref{AlgC}). 
We can also express the conserved quantities in terms of $\bf{A}^{-1}$,  since $\sum_{i',j'} A^{-1}_{ii'} A^{-1}_{jj'} Q_{i'j'} = \bar{Q}_{ij}$ with
\begin{equation}
\bar{Q}_{ij} = c^\dag_i ({\bf A}^{-1} \bm{c})_j \pm c^\dag_j ({\bf A}^{-1} \bm{c})_i,
\label{Qbar}
\end{equation}
in agreement with Eq.\ \eqref{Statebis} (despite the latter holds only for fermions, Eq. \eqref{Qbar}  remains valid also for bosons). 

In the natural sp basis in which $A^{\dagger}$ has the form \eqref{Schmidt}, Eq.\ \eqref{Qconserved} leads to 
\begin{subequations}
\begin{eqnarray}
Q_{kl} &=& \sigma_k a_{\bar{k}}^\dagger a_l+\sigma_l a_{\bar{l}}^\dagger a_k,\ k\leq l,\label{20a}\\
Q_{\bar{k}\bar{l}} &=& \sigma_k a_k^{\dagger}a_{\bar{l}}+\sigma_l a_l^\dagger a_{\bar{k}},\ k\leq l,\label{2-b}\label{20b}\\
Q_{\bar{k}l} &=& \sigma_k a_k^\dagger a_l-\sigma_l a_{\bar{l}}^\dagger a_{\bar{k}},
\label{20c}
\end{eqnarray}
\label{Qconservedf}
\end{subequations}
for fermions and 
\begin{equation}
Q_{kl}=\sigma_k b_k^\dagger b_l-\sigma_l b_l^\dagger b_k\,,\;\;k<l\,,
\label{Qconservedb}
\end{equation}
for bosons. These ``normal'' conserved operators satisfy  $SU(2)$ algebras for each pair $k,l$ if properly scaled (Eqs.\ \eqref{B8}--\eqref{B12}), and are then angular momentum-like operators, with similar  eigenvalues (see App.\ \ref{ApB}). 

In the fermion case,  the $\frac{3}{2}n$ conserved  operators $Q_{kk}\propto a_{\bar k}^\dag a_k$, $Q_{\bar k\bar k}\propto a^\dag_ka_{\bar k}$ and $Q_{\bar k k}\propto\frac{1}{2}(a_k^\dag a_k-a_{\bar k}^\dag a_{\bar k})$, do not depend on the $\sigma_k$ except for a constant and  also satisfy $SU(2)$ commutation relations (Eq.\ \eqref{B11}). They  are conserved for   general ``paired'' states of the form 
\begin{equation}
\!\!|\psi_m\rangle=\!\!
\sum_{m_1,\ldots, m_d}\!\!\!\!\Gamma_{m_1\ldots m_d}(a_1^\dag a_{\bar{1}}^\dag)^{m_1} \!\ldots (a_{d}^\dagger a_{\bar{d}}^\dag)^{m_d}|0\rangle\,,
\label{Pairing}
\end{equation}
for $m_k=0,1$, $d=\frac{n}{2}$ and $\sum_{k}^d m_k=m$, with \eqref{State} recovered for $\Gamma_{m_1\ldots m_d}\propto\sigma_{1}^{m_1}\ldots\sigma_{d}^{m_d}$.  Hence, the additional  $4\binom{n/2}{2}$ conserved quantities \eqref{20a}--\eqref{20b} for $k<l$ and  \eqref{20c} for $k\neq l$ are those that distinguish the state \eqref{State} from \eqref{Pairing}. 

If we consider the bosonic version of the state \eqref{Pairing} ($a^\dag_{k,\bar k}\rightarrow b^\dag_{k,\bar k}$, $m_k=0,1,2,\ldots$), it has in general just $\hat N$ and the $n/2$ operators $Q_{\bar k k}=\frac{1}{2}(b^\dag_k b_k-b^\dag_{\bar k} b_{\bar k})$ conserved. A bosonic ``paired'' pair condensate, arising when the $\sigma_k$ in \eqref{Schmidtbos} come in degenerate pairs $\sigma_k=\sigma_{\bar k}$ (as $(b^\dag_k)^2+(b^\dag_{\bar k})^2=2\tilde b^\dag_k \tilde b^\dag_{\bar k}$ for $\tilde b^\dag_{k,\bar k}=\frac{1}{\sqrt{2}}(b^\dag_k\pm ib^\dag_{\bar k})$), has 
 the additional $4\binom{n/2}{2}$ conserved  operators $Q_{kl}$,   $Q_{\bar k \bar l}$ for $k<l$ and $Q_{\bar k l}$ for $k\neq l$, 
 defined as in \eqref{Qconservedf}
  with $a,a^\dag\rightarrow b,b^\dag$ and $+\rightarrow -$ in \eqref{20a}--\eqref{20b},   satisfying \eqref{19},   which lead again to Eq.\ \eqref{L}.

On the other hand,  a typical random state $|\psi\rangle$ of $2m$ particles with $m\geq 2$ (and $m\leq n/2-2$ for fermions) has no conserved one-body operators  satisfying $Q|\psi\rangle=\lambda|\psi\rangle$  other than  the particle number, such that the nullspace of ${\bf C}^{(1,1)}$ has  just dimension $1$. 

The rather high dimensionality of the  nullspace of ${\bf C}^{(1,1)}$ in the state \eqref{State} suggests that these states are very special.  In fact, excluding as always empty levels for bosons and both empty  and fully occupied levels for fermions, we can claim (see also App.\ \ref{ApC}) the following conjecture for $m\geq 1$ (and $m\leq n/2-1$ for fermions):

{\bf Conjecture:}. 
{\it Amongst $2m$-particle states with support on an n-dimensional sp space  having a full rank one-body DM   $\bm\rho^{(1)}$, and $\mathbbm{1}-\bm\rho^{(1)}$ also full rank for fermions,  the states \eqref{State} have the maximum number of conserved one-body operators}.  

On the other hand, regarding conserved pair creation or annihilation operators, i.e., linear in $c_j c_i$ or $c^\dag_i c^\dag_j$, which are determined by the covariance matrices 
\eqref{C20}-\eqref{C02}, we can demonstrate (see App.\ \ref{ApD}): 

{\bf Proposition 2}. {\it For $m\geq 2$ (and $m\leq n/2-2$ for fermions), the state \eqref{State} has no conserved operators linear in $c_j c_i$ or $c^\dag_i c^\dag_j$.}  

This result is remarkable, since for $m=1$, there are obviously $\frac{n(n\mp 1)}{2}-1$ linearly independent  pair annihilation operators $A_\mu=\sum_{i,j}A^*_{\mu\,ij}c_j c_i$ satisfying $A_\mu A^\dag|0\rangle=0$ (i.e., those $A^\dag_\mu$  creating orthogonal pair states 
such that $\langle 0|A_\mu A^\dag|0\rangle=0$). None of them survives strictly for $m \geq 2$, a result which is connected with the non-singularity of the two-body DM $\bm\rho^{(2)}$ in any state \eqref{State} for $m\geq 2$ (even though its lowest eigenvalue may be small, it is nonzero, see App.\ \ref{ApD}). This result exposes the fact that the pair condensate is not a strict bosonic condensate for $m\geq 2$. For fermions, a similar result holds for pair creation operators due the particle-hole symmetry: Even though for $m=n/2-1$   
the state \eqref{State} has obviously the same number of conserved pair creation operators (those $\bar{A}^\dag_\mu$  orthogonal to $\bar{A}$, such that $\bar{A}^\dag_\mu \bar{A}|\bar{0}\rangle=0$), they are not conserved for $m\leq n/2-2$. On the other hand, for bosons the 
matrix \eqref{C02} is positive definite and hence there is no conserved pair creation operator if $m\geq 2$. 

We also notice that for recognizing the conserved operators $Q_{ij}$, it is sufficient to consider the matrix
\begin{equation}
\rho^{(1,1)}_{ij,i'j'} = \langle c_j^\dag c_i c_{i'}^\dag c_{j'} \rangle\,,
\label{rho11}
\end{equation}
instead of \eqref{C11}, since $\langle Q_{\alpha}\rangle=0$ and $ \langle Q_\alpha^\dag Q_\alpha \rangle = 0$ iff $Q_\alpha |\psi\rangle = 0$. Hence we can claim that Eq.\ \eqref{rho11} has $L_n-1$ null eigenvalues iff the state has the form \eqref{State}. 
This matrix has a fixed trace for definite particle number states: $\text{Tr}[{\bm \rho}^{(1,1)}] = N(n\mp(N-1))$. Its nullspace  directly determines those conserved quantities satisfying $Q_\alpha |\psi\rangle = 0$.  

A final comment is that in the bosonic case, for $A^\dag = A^\dag_0$ the uniform pair creation operator, $Q_{kl}=Q^0_{kl} \propto x_k p_l - p_l x_k$ is the angular momentum associated with the $k,l$ plane (see App.\ \ref{ApB}),  with $x_k = \frac{b_k+b^\dag_k}{\sqrt{2}}$, $p_k =  \frac{b_k- b_k^\dag}{\sqrt{2}i}$ the associated coordinate-momentum operators. Thus, 
$Q^0_{kl}|\psi\rangle = 0$ $\forall\,k<l$ iff $\psi(\bm{x}) = \langle \bm{x}|\psi\rangle \equiv \psi(r)$,  with $r=\sqrt{\sum_i x_i^2}$. If in addition the state has definite particle number, i.e.\ is of the form \eqref{State}, these functions $\psi(r)$ are then the isotropic eigenfunctions of the isotropic $n$-dimensional harmonic oscillator $\psi_{2m,0,0}(r)$. In the general case the conserved quantities and state can also be obtained from the latter via Eqs.\  \eqref{trans}--\eqref{transQ}. 

\subsection{Hamiltonians and operators having the pair condensate as exact eigenstate}

We are now in a position to determine the most general two-body Hamiltonian $H=h+V$, with $h=\sum_{i,j} h_{ij} c_i^\dag c_j$ and $V=\frac{1}{4} \sum_{i,j,k,l} V_{ij,i'j'} c_i^\dag c_j^\dag c_{j'} c_{i'}$, having the pair condensate $|m\rangle_2$ as exact eigenstate,  
\begin{equation}
H |m\rangle_2 = \lambda_m |m \rangle_2.
\label{eigen}
\end{equation}
Since $\tilde{Q}_{ij} = Q_{ij} - \langle Q_{ij} \rangle = Q_{ij}$ and $\tilde{N} = \hat{N} - \langle \hat{N} \rangle = 0$ within a subspace with definite particle number, Eq. \eqref{Hamiltonian} leads to the following hermitian Hamiltonian
\begin{equation}
H_Q = \tfrac{1}{8} \sum_{i,j,i',j'} V_{ij,i'j'} Q_{ij}^\dag Q_{i'j'}\,,
\label{Hambis}
\end{equation}
which satisfies Eq.\ \eqref{eigen} with $\lambda_m=0$ $\forall\,m$. We used the evident symmetry $Q_{ij}=\pm Q_{ji}$ ($+$ fermions, $-$ bosons) and summed over all $i,j$, assuming $V_{ij,i'j'}=\pm V_{ji,i'j'}=\pm V_{ij,j'i'}=V_{i'j',ij}^*$ (for $H_Q$ hermitian).  Furthermore, if the matrix $V_{\alpha\beta}\equiv V_{ij,i'j'}$  is positive definite, $H_Q$ is positive semidefinite and hence \eqref{State} is the GS of \eqref{Hambis}, being also non-degenerate within the subspace of fixed particle number, since the $Q_{ij}$ define the state univocally. 

Moreover, Eq.\ \eqref{Hamiltonianp} leads to the general conserved two-body operator 
\begin{equation}
H'_Q=\sum_{i,j}h_{ij}Q_{ij}+V_{\mu,ij}O_\mu Q_{ij}\,,
\label{H2g}    
\end{equation}
 where $O_\mu$ are arbitrary one-body operators.    

Therefore, we can claim the following important theorem which is proved in detail in App.\ \ref{ApE}. \\
{\bf Theorem 2}. {\it Within the subspace of $2m$-particle states, with $m\geq 2$ (and $m\leq n/2-2$ for fermions) the most general two-body operator having \eqref{State} as exact eigenstate (except for constants  or terms $\propto$ $\hat N$ or $\hat N^2$) is given by Eq.\ \eqref{H2g}, which satisfies $H'_Q|m\rangle_2=0$.}
 
In particular the most general hermitian two-body Hamiltonian having \eqref{State} as eigenstate is obtained from \eqref{H2g} imposing hermiticity, 
i.e.\ setting $V_{\mu,ij}O_{\mu}\rightarrow V_{i'j',ij}Q_{i'j'}^\dag$ as in \eqref{Hambis}, with $V_{i'j',ij}$ hermitian, and restricting the one-body part to hermitian combinations.  

Previous considerations hold for any sp basis. 
In the natural sp basis,
$Q_{kk}+Q_{\bar k\bar k}$, 
$i(Q_{kk}-Q_{\bar k\bar k})$ and $Q_{\bar k k}$ are hermitian for fermions and can be included in \eqref{H2g} through the one-body term. In addition, if  $\sigma_k=\sigma_l$ for some pair $k,l$, 
$Q_{kl}^\dag$ (as well as  $Q_{\bar k l}^
\dag$ and $Q_{\bar k \bar l}^\dag$ for fermions) becomes proportional to another operator $Q_{kl}$ of this set, and hence is also conserved, implying that extra  hermitian conserved one body terms $\propto Q_{kl} + Q_{kl}^\dag$ or $i(Q_{kl}-Q_{kl}^\dag)$ can be added to the Hamiltonian. 

In particular, for fermions in the $a_k,a_{\bar{k}}$ basis and   $V_{\alpha \beta} = V_{\alpha} \delta_{\alpha \beta}$, with   $V_{kl} = V_{\bar{k}l} = V_{k\bar{l}} = V_{\bar{k}\bar{l}}$, Eq.\ \eqref{Hambis} becomes  
\begin{subequations}
\label{Hambis2}
\begin{equation}
    \begin{split}
&\!\!\!\!\!H_Q^F\!=\! \sum_k[\epsilon_k \hat{n}_k + \tfrac{3}{4}  V_{kk} \sigma_k^2 (a_k^\dag a_k - a_{\bar{k}}^\dag a_{\bar{k}})^2]  
\\
&\!\!\!\!\!-\tfrac{1}{2} \sum_{k\neq l} V_{kl}  
[\sigma_k\sigma_l( A_k^\dag A_l \!+\!  A_l^\dag  A_k) \!+\! (\sigma_k^2\!+\!\sigma_l^2)\hat{n}_k\hat{n}_l],
\label{Hambis2F}
\end{split}
\end{equation}
where $\hat{n}_k =\frac{1}{2}(a^\dag_k a_k+ a^\dag_{\bar{k}} a_{\bar{k}})$, $A_k^\dag = a_k^\dag a_{\bar{k}}^\dag$  
and $\epsilon_k = \sum_{l \neq k} V_{kl} \sigma_l^2$. This is the most general two-body pairing-type Hamiltonian having \eqref{State} as eigenstate with null eigenvalue, and as a GS  if all $V_{kl}$ are positive (sufficient condition).  We remark that only in the special case $V_{kl}=\frac{\varepsilon_k-\varepsilon_l}{\sigma_k^2-\sigma_l^2}$  (with $\varepsilon_k$ arbitrary parameters), the Hamiltonian \eqref{Hambis2F} reduces to those of \cite{DES.01,DPS.04,LDO.19} (see also \cite{Rich0,Rich1,DDEP.04}), 
which are exactly solvable (for all  eigenstates). 

Similarly, for bosons in the  $b_k^\dag$ basis (and setting again $V_{\alpha\beta} = V_\alpha\delta_{\alpha\beta}$), the Hamiltonian \eqref{Hambis} leads to 
\begin{equation}
\begin{split}
H_Q^B &\!=\! \tfrac{1}{2}\sum_k \epsilon_k \hat{n}_k  \\
&\!\!\!\!\!-\! \tfrac{1}{4} \sum_{k\neq l} V_{kl}
[\sigma_k\sigma_l(b^{\dag 2}_k b^2_l\!+\!b_l^{\dag 2}b_k^2) \!-\!(\sigma_k^2\!+\!\sigma_l^2)\hat{n}_k\hat{n}_l], 
\label{Hambis2B}
\end{split}
\end{equation}
\end{subequations}
where $\hat{n}_k = b_k^\dag b_k$ and $\epsilon_k = \sum_{l\neq k} V_{kl} \sigma_l^2$. In the pairing case, where the $\sigma_k$ come in degenerate pairs $\sigma_k=\sigma_{\bar k}$, Eq.\ \eqref{Hambis2B} becomes similar to \eqref{Hambis2F} after a trivial sp transformation, and reduces again  to those of \cite{DES.01,DPS.04,LDO.19} for the previous choice of $V_{kl}$.    

In the special case $V_{\alpha\beta} = \frac{1}{2} \delta_{\alpha \beta}$, i.e.\ $V_{kl}=1$ in \eqref{Hambis2F}--\eqref{Hambis2B}, these two  Hamiltonians acquire the simple form
\begin{equation}
H_A = \tfrac{1}{4} \sum_{i,j} Q_{ij}^\dag Q_{ij}=\hat M-\hat M_A\,,
\label{Hambis3}
\end{equation}
where $\hat{M} =\hat{N}/2$ is the pair number operator 
and 
\begin{equation}
\hat M_A
= A^\dag A - \tfrac{1}{2}(\hat{M}-1) ([A,A^\dag]-1)\,, \label{ncobd}
\end{equation}
for both fermions and bosons. As $H_A$ is positive semidefinite and $H_A |m\rangle_2 = 0\,\forall\,m$, the operator $\hat M_A$ satisfies 
\begin{equation}
\hat M_A|m\rangle_2=
m|m\rangle_2\,, 
\label{ncob}
\end{equation}
with $m$ its {\it largest} eigenvalue. Hence $\hat M_A$ behaves  as a {\it pair number operator} for pair condensates $|m\rangle_2$ built with the operator $A^\dag$.  

If $A,A^\dag$ are replaced by  standard boson operators $b,b^\dag$, the r.h.s.\ in   \eqref{ncobd} reduces to $b^\dag b=\hat N_b$,  satisfying 
$\hat N_b|m\rangle_1=m|m\rangle_1$ for the standard condensates \eqref{Statem1}. Eq.\ \eqref{Hambis3}
is thus an extension to the pair regime of previous  Hamiltonian \eqref{Hb}. 
 Nonetheless, while $\hat M_A$   has a set of integer eigenvalues $m$ with the condensates $|m\rangle_2$ as exact eigenstates, it also has other noninteger eigenvalues, {\it smaller} than $m=N/2$ within each fixed $N$  subspace, as $H_A$ in Eq.\ \eqref{ncobd} is positive semidefinite. 
Besides, as the nullspace of $H_A$ is spanned just by the set of condensates $|m\rangle_2$ with $m$ integer,  $H_A>0$ (hence $\hat M_A<N/2$) in any {\it odd}-particle number subspace.  

If instead of \eqref{Qconservedf}-\eqref{Qconservedb}  one uses in \eqref{Hambis3} the conserved operators \eqref{Qbar}, we obtain a positive semidefinite Hamiltonian expressed in terms of the dual   operators $\bar{A}^\dag$, $\bar A$ (Eq.\ \eqref{Abar}, here assumed normalized: $\langle 0|\bar{A}\bar{A}^\dag|0\rangle=1$), given  by 
\begin{equation}
\begin{split}
\bar H_{\bar A} &= \tfrac{1}{4} \sum_{i,j} \bar{Q}_{ij}^\dag \bar{Q}_{ij}\\ &=
\tfrac{1}{2}(\hat{M}\mp\tfrac{n}{2}-1)([\bar{A},\bar{A}^{\dagger}]-1)-\bar{A}^{\dagger}\bar{A}\,,\label{Hbar}
\end{split}
\end{equation} 
which also has the same previous condensates $|m\rangle_2\propto (A^\dag)^m|0\rangle$ as GS with null eigenvalue: $\bar H_{\bar A}|m\rangle_2=0$ $\forall\,m$. 
We finally mention that while  several effective bosonised Hamiltonians have been employed in relation with coboson approaches (see discussions in e.g.\ \cite{CB.02,CBD.08,CSC.16}), no bosonic approximations to  $A^\dag$ or other  operators have been invoked in  Hamiltonians \eqref{Hambis}, \eqref{Hambis2}, \eqref{Hambis3} and \eqref{Hbar} for having the pair condensate as exact  GS.  

\subsection{Exact condition for pair condensation}
Projecting Eq.\ \eqref{ncob} onto $_2\langle m|$ and using $_2\langle m|m\rangle_2=1=\frac{1}{2}\sum_{i,j}|A_{ij}|^{2}$, we arrive at a quadratic matrix equation of the form $\frac{1}{2}\bm{A}^{\dagger}{\bf H}_m\bm{A}=0$, with $\bm{A}$ a vector of elements  $A_{ij}$ ($=\mp A_{ji}$) and ${\bf H}_m$ an $\bm A$-independent matrix, determined by one-and two-body averages:
\begin{equation}
{\bf H}_m=m\mathbbm 1-\tfrac{1}{2}\tilde{\bm  \rho}^{(2)}_m\,,
\label{Hm}
\end{equation}
where 
\begin{subequations}
\begin{eqnarray}
\!\!\!\!\!\!\!\tilde{\bm \rho}^{(2)}_m&=&\bm\rho^{(2)} \pm \tfrac{1}{2}(m-1)(\mathbbm{1}\otimes_s\bm\rho^{(1)}+\bm\rho^{(1)}\otimes_s\mathbbm{1})\;\;\;\;\;\;\label{Lam21}\\&=&\tfrac{1}{2}[(1+m)\bm{\rho}^{(2)}+(1-m)(\bar{\bm{\rho}}^{(2)}-{\mathbbm 1}\otimes_s\mathbbm{1})],
\label{Lam22}
\end{eqnarray}
\label{Lam2}
\end{subequations}
$\!\!\!$with ${\bm\rho}^{(2)}$,  $\bar{\bm\rho}^{(2)}$ defined as in \eqref{C20}--\eqref{C02},   $(A\otimes_s  B)_{ij,kl}=A_{ik}B_{jl}\mp A_{il}B_{jk}$ the antisymmetrized (symmetrized) product for fermions (bosons) and $\mathbbm{1}_{ij}=\delta_{ij}$.   Using again that \eqref{Hambis3} is positive semidefinite, the
matrix ${\bf H}_m$ should also be  positive semidefinite (within the antisymmetric or symmetric subspace) so that ${\bm A}^\dag {\bf H}_m {\bm A}=0$ implies ${\bf H}_m\bm A=\bm 0$, 
which leads to 
\begin{subequations}
\begin{equation}
\tfrac{1}{2} \tilde{\bm \rho}^{(2)}_m \bm A=m\bm A\,,
\label{Eq1}
\end{equation}
or equivalently,
\begin{equation}
\tfrac{1}{2}[(1+m)\bm{\rho}^{(2)}+(1-m)\bar{\bm{\rho}}^{(2)}]\bm A=(1+m)\bm A\,.\label{Eq2}
\end{equation}
\label{Eqq}
\end{subequations}
Explicitly, these equations imply (for $A_{ij}=\mp A_{ji}$) 
\begin{subequations}
\begin{equation}
\!\!\!\!\tfrac{1}{2} \sum_{k,l}[\rho_{ij,kl}^{(2)} \pm (m-1)( \delta_{ik}\rho^{(1)}_{jl}+\rho^{(1)}_{ik} \delta_{jl})]A_{kl}=mA_{ij}\,,
\label{Condition1}
\end{equation}
or equivalently 
\begin{equation}
\tfrac{1}{2}\sum_{k,l}[(1+m)\rho_{ij,kl}^{(2)}+(1-m)\bar{\rho}_{ij,kl}^{(2)}]A_{kl}=(1+m)A_{ij}.
\label{Condition2}
\end{equation}
\label{Cnd}
\end{subequations}
Therefore, we can claim the following theorem:\\
{\bf Theorem 3}. {\it  An  $N=2m$ particle state (fermionic or bosonic) is a pair condensate of  the form (\ref{State})
iff the largest  eigenvalue of the associated matrix  $\frac{1}{2}\tilde{\bm \rho}^{(2)}_m$,   with $\tilde{\bm \rho}^{(2)}_m$ given by \eqref{Lam2},  has the integer value $m$ (Eq.\  \eqref{Eqq}). In this case  the corresponding  eigenvector $\bm A$ (normalized as $\bm A^\dag\bm A=2$) is just the vector of elements $A_{ij}$ (not depending on $m$) determining the normalized pair creation operator $A^\dag$ of the condensate.}

Hence, with $\tilde{\bm \rho}^{(2)}_m$  
we can exactly detect, through its maximum eigenvalue, if a $2m$-particle pure state is a coboson condensate, in which case we can recover it  completely through  the associated eigenvector. 
This result holds for both fermions and bosons. 
 
In contrast, such state cannot be fully recognized through the one-body DM $\bm\rho^{(1)}$, which just has maximum rank but no other special feature. And while in the state \eqref{State} the two-body DM $\frac{1}{2}\bm\rho^{(2)}$ has always a maximum eigenvalue $\lambda^{(2)}_{\rm max}\geq 1$ for fermions and $\geq m$ for bosons \cite{CR.24} 
\footnote{The factor  $\frac{1}{2}$ in $\bm\rho^{(2)}$ applies when labels $ij$,  $i'j'$ in $\rho^{(2)}_{ij,i'j'}$ are  arbitrary as here assumed,  but is to be omitted when  restricted to independent  pairs ($i<j$ for fermions, $i\leq j$ for bosons, with  $b_i^2\rightarrow \frac{1}{\sqrt{2}}b_i^2$), as in \cite{CR.24} and section  \ref{III}}, this also occurs in other states. 
 
As a check, for a general two-particle state $|\Psi\rangle=A^\dag|0\rangle$ ($m=1$),  $\tilde{\bm\rho}^{(2)}_m=\bm\rho^{(2)}$, with $\bm\rho^{(2)}=\bm A \bm  A^\dag$ for fermions and bosons (i.e., $\rho^{(2)}_{ij,kl}=A_{ij}A^*_{kl}$),  normalization implying ${\bm A}^\dag \bm A=2$. Then  Eq.\ \eqref{Eq1} is always fulfilled. Similar arguments hold for $m=n/2-1$ for fermions. And for a standard $N=2m$ boson condensate ($A^\dag\propto {b_1^\dag}^2$), just  $\rho^{(1)}_{11}=2m$,  $\rho^{(2)}_{11,11}=2m(2m-1)$ and $A_{11}$ are nonzero (in the natural sp basis), leading again to   Eq.\ \eqref{Eq1}.  

In the fermionic case any $2m$-particle SD leads as well to an eigenvalue $m$ of $\frac{1}{2}\tilde{\bm\rho}^{(2)}_m$, since they can be written as $(A^\dag)^m|0\rangle\propto\prod_{k=1}^m c^\dag_k c^\dag_{\bar k}|0\rangle$
for $\bm A$ of rank $2m$ (just $\sigma_1,\ldots,\sigma_m$ are nonzero). Nonetheless,  this eigenvalue becomes $\binom{2m}{2}$-fold degenerate,  as in this case 
$\bm\rho^{(1)}=\Pi_{2m}$, $\bm\rho^{(2)}=\Pi_{2m}\otimes_s\Pi_{2m}$, with $\Pi_{2m}$  the projector onto the occupied sp space, so that it can be  distinguished from a ``true'' full rank condensate through its degeneracy. 

Similarly, a  state 
$|\Psi\rangle\propto (\prod_{k=1}^lc^\dag_k c^\dag_{\bar k})({A'}^\dag)^{m-l}|0\rangle$ with $m>l$ and rank $\bm A'>2m-2l$,  also leads to an eigenvalue $m$ for fermions with degeneracy $\binom{2l}{2}$, since it is 
the limit of the normalized condensate $\propto(\sum_{k=1}^l c^\dag_kc^\dag_{\bar k}+\varepsilon {A'}^\dag)^m|0\rangle$ for $\varepsilon\rightarrow 0$ (here ${A'}^\dag$ denotes  a pair creation operator in the sp space orthogonal to the $k,\bar k$). 

On the other hand, we remark that the present method is exact and its validity does not depend on the extent of bosonic properties displayed by the pair created by $A^\dag$, which is related to its entanglement \cite{COW.10,CKL.05}, 
nor to the presence of off-diagonal long range order \cite{Yang.62}. 

{\it Odd states}. Finally, for fermions, we can also recognize states with an odd particle number of the form 
\begin{equation}
|\Psi_{\rm odd}\rangle \propto c_i^\dag (A^{\dag})^{m}|0\rangle,
\label{Stateodd}
\end{equation}
obtained by creating an arbitrary sp state on  the condensate \eqref{State}. For such states, the one body DM has an eigenvalue equal to $1$, corresponding to $c^\dag_i c_i$,  since \eqref{Stateodd}
is equivalent to  $c^\dag_i ({A'}^\dag)^m|0\rangle$, 
with ${A'}^\dag$ obtained by removing sp state $i$ from $\bm A$ and having then rank $n-2$. This  also leads to a zero eigenvalue of $\bm\rho^{(1)}$ associated to some sp state $\bar i$ orthogonal to $i$ and the sp space occupied by ${A'}^\dag$. Thus, $\frac{1}{2}\tilde{\bm\rho}^{(2)}_m$ is split in two blocks (one comprising sp states $i,\bar i$ and the other the orthogonal subspace), having also an eigenvalue $m$, corresponding to the second block. Then we can reconstruct ${A'}^\dag$ with the corresponding eigenvector. Similar considerations hold for states $c_i(A^\dag)^m|0\rangle$, as they are equal to $m[c_i,A^\dag] (A^{\dag})^{m-1}|0\rangle$ and $[c_i,A^\dag]$ is a sp creation operator.\

\subsection{Proximity to a pair condensate}
When $\bm \rho^{(1)}$ and $\bm\rho^{(2)}$ are determined by an arbitrary $2m$-particle normalized state  $|\Psi\rangle$, the matrix \eqref{Hm} satisfies 
\begin{equation}
\tfrac{1}{2} {\bm A}^\dag {\bf H}_m \bm A=\langle \Psi|H_A|\Psi\rangle\,,
\label{APA}
\end{equation}
for any vector $\bm A$ of elements $A_{ij}$ ($=\mp A_{ji}$),  with $H_A$ the Hamiltonian \eqref{Hambis3} for the corresponding pair creation operator $A^\dag$.  
 Eq.\ \eqref{APA} also holds for general $2m$-particle mixed states $\hat \rho$, replacing $\langle\Psi|\ldots|\Psi\rangle\rightarrow {\rm Tr}[\hat\rho\ldots]$. 
As $H_A$ is positive semidefinite, 
 ${\bm A}^\dag {\bf H}_m {\bm A}\geq 0$, vanishing iff $|\Psi\rangle$ is the $m$ pair condensate $|m\rangle_2\propto (A^\dag)^m|0\rangle$ associated to $\bm A$   (or in general iff $\hat\rho=|m\rangle_2\langle m|$), 
according to Theorem 3.

For a $2m$-particle state $|\Psi\rangle$, the quantity  
\begin{equation}
D_2(|\Psi\rangle)=m-\tfrac{1}{2}\lambda_{\rm max}(\tilde{\bm\rho}^{(2)}_m)
\label{DE}\,,
\end{equation} 
where   $\lambda_{\rm max}$ denotes the largest eigenvalue of the  $\tilde{\bm\rho}_m^{(2)}$ determined by $|\Psi\rangle$, can be considered as a simple measure of the {\it proximity} of  $|\Psi\rangle$ to an $m$-pair condensate:  From Theorem 3 and Eq.\ \eqref{APA} it follows that $D_2$ satisfies: 

1) $D_2(|\Psi\rangle)\geq 0$, with $D_2(|\Psi\rangle)=0$ iff $|\Psi\rangle$ is an $m$-pair condensate 
(including the limit cases discussed before).

2) \vspace*{-.75cm}

\begin{subequations}
\begin{align}    D_2(|\Psi\rangle)&=\langle\Psi|H_A|\Psi\rangle\label{D1}\\
&=\underset{A'}{\rm Min}
\langle \Psi|H_{A'}|\Psi\rangle\,,
\label{D2}
\end{align}
\end{subequations}
where $H_A$ is the Hamiltonian \eqref{Hambis3}  determined by the associated eigenvector $\bm A$ ($\frac{1}{2}\tilde{\bm\rho}_m^{(2)}\bm A=\lambda_{\rm max}\bm A$, with $\bm A^\dag \bm A=2$) and ${A'}^\dag$ any other normalized pair creation operator.  Eq.\ \eqref{D1} follows from \eqref{Hm}--\eqref{APA} since by Eq.\ \eqref{D1}, $D_2(|\Psi\rangle)=\frac{1}{2}\bm A^\dag {\bf H}_m\bm A$, 
while  $\frac{1}{2}{{\bm A}}^\dag {\bf H}_m\bm A\leq \frac{1}{2}{{\bm A}'}^\dag {\bf H}_m\bm A'=\langle\Psi|H_{A'}|\Psi\rangle$  for any $\bm A'$ with the same normalization, since  $m-\frac{1}{2}\lambda_{\rm max}$ is the lowest eigenvalue of ${\bf H}_m$. 

Thus, the condensate  $|m\rangle_2\propto(A^\dag)^m
|0\rangle$ obtained from the  eigenvector $\bm A$ associated to $\lambda_{\rm max}$, satisfying $H_{A}|m\rangle_2=0$ and hence minimizing $\langle H_{A}\rangle$ among $2m$-particle states,  provides an $m$-pair approximation to $|\Psi\rangle$, which is  ``optimum'' in the sense that $\langle \Psi|H_A|\Psi\rangle$ is minimum (Eq.\ \eqref{D2}), i.e., closest to $0$.   This minimum is $0$ iff $|\Psi\rangle$ is an $m$ pair condensate.  Moreover, for ``true''  $m$-pair condensates (i.e., excluding SDs and related limit cases in the fermionic case)  the minmum in Eq.\ \eqref{D2} is unique, as the maximum eigenvalue $\lambda_{\rm max}$ is nondegenerate. 

Notice that an analogous measure for the proximity to  a standard $m$-particle condensate among $m$-particle states would be 
$D_1(|\Psi\rangle)=m-\lambda_{\rm max}(\bm\rho^{(1)})$, which coincides with $\langle \Psi|H_{b}|\Psi\rangle$ 
 for $H_b$ given by  
\eqref{Hb} and $b$ the eigenvector  associated to the maximum eigenvalue of the one-body DM $\bm\rho^{(1)}$. 

\subsection{Generalization}
Let us now consider the states \eqref{State2}--\eqref{State2m}, involving coherent or statistical mixtures of condensates $|m\rangle_2$. All these  states have obviously definite number parity (even) yet not definite particle number. 

In first place, since $Q_{ij}|m\rangle_2=0 \ \forall \ m$, all previous operators \eqref{Qconserved} will also be conserved in any of these states i.e., $Q_{ij}|\Psi_A\rangle=0$, $Q_{ij}\rho_A=0$. On the other hand, the number operator $\hat N$ is no longer conserved, so that in general, $L_n\rightarrow L_n-1$ in Eq.\ \eqref{L}. Then, the general Hamiltonian 
\eqref{Hambis} will still satisfy 
\begin{equation}
H_Q |\Psi_A\rangle =0
\end{equation}
and also $H_Q\,\rho_A=0$, for any $f$ and $\alpha_{mm'}$ respectively. Thus, $H_Q$ will have  \eqref{State2} as a (degenerate) GS if $V_{ij,i'j'}$ is positive definite. In particular, the same holds for the Hamiltonians \eqref{Hambis2}--\eqref{Hambis3}. 

Regarding Eqs.\ \eqref{Condition1}--\eqref{Condition2}, they can be easily generalized introducing $m$ as $\hat{M}=\hat N/2$ within the mean values, such that they become 
\begin{equation}
\tfrac{1}{2} \sum_{k,l}[\rho_{ij,kl}^{(2)} \pm(
\tilde\rho^{(1)}_{ik}\delta_{jl} + \delta_{ik}\tilde\rho^{(1)}_{jl})]A_{kl}=\langle \hat M\rangle A_{ij}.
\label{Condition2f}
\end{equation} 
where $\tilde{\bm\rho}^{(1)}$ is a weighted average of one-body DMs for each $m$:
\begin{equation}
\tilde\rho^{(1)}_{ij} = \langle (\hat{M}-1)c_j^\dag c_i\rangle.
\label{rt}
\end{equation}
Hence, we obtain: \\
{\bf Theorem 4.} {\it A state is of the form \eqref{State2}  or in general \eqref{State2m},  iff the matrix on the l.h.s.\ of \eqref{Condition2f} has a maximum eigenvalue equal to $\langle \hat M\rangle$, where $\langle \hat M\rangle=\frac{1}{2}{\rm Tr}\,\bm\rho^{(1)}=\frac{1}{2}\langle \hat N\rangle$ is the  average pair  number. In this case the corresponding eigenvector is the vector $\bm A$.}
 
Thus, in order to identify any of such states, one should compute the maximum eigenvalue of this matrix and compare it with the average pair number. Of course, since these equations are based on number conserving averages, this test will not distinguish between the states \eqref{State2}--\eqref{State2m}, since $\langle \hat M\rangle$, $\bm\rho^{(2)}$ and $\bm\rho^{(1)}$ just depend on  $p_m=\alpha_{mm}$. Additional information on average pair creation $\langle c^\dag_i c^\dag_j\rangle$ or annihilation operators should obviously be incorporated to distinguish between these states. And further state tomography is required for obtaining the  $p_m$'s. Nonetheless, the pair creation operator $A$ is still exactly obtained from the corresponding eigenvector $\propto \bm A$  of this matrix. 

We also remark  that in the case of an odd number-parity state, its maximum eigenvalue will not reach $\langle \hat M\rangle$. Hence, nor will it reach $\langle \hat M\rangle$ in any mixture containing odd particle number states. 

\section{Illustrative results\label{III}}

We now show  typical results for the exact GS of model Hamiltonians with attractive pairing-type couplings, in both bosonic and fermionic systems. Their GS will be of the  general paired form \eqref{Pairing} (and its extension to the bosonic case), and can then be expected to be close to a pair condensate at least in some limit.  Model Hamiltonians of this type have been extensively used  in nuclear and condensed matter physics (see e.g.\ \cite{RS.80,LWS.80,DHJ.03,bcs.57,DR.01,RCR.98,DGR.18}, including  bosonic models \cite{DS.01}).  The exact results were obtained through direct numerical diagonalization.

\subsection{Bosonic system}
In the bosonic case we  consider the Hamiltonian 
\begin{equation}
H_B=\sum_k\varepsilon_k b^\dag_kb_k -gA^\dag A\label{HBg}\,,
\end{equation}
where $A^\dag=\frac{1}{\sqrt{2}}\sum_k \sigma_k (b^\dag_k)^2$, $\sum_k \sigma_k^2=1$ and  $k=1,\ldots,n$. As stated below Eq.\ \eqref{Pairing}, if  the $\sigma_k$ come in degenerate pairs $\sigma_k=\sigma_{\bar k}$,   $A^\dag$ can be rewritten as $\sqrt{2}\sum_{k=1}^{n/2}\sigma_k \tilde b^\dag_{k}\tilde b^\dag_{\bar k}$, and the coupling in \eqref{HBg} acquires the standard  pairing form  involving pair creation in  conjugate sp states  $k,\bar k$. In the general case the meaning is similar except that pairs are created in the same sp state. For $g>0$ the GS will   prefer to maximize $A^\dag A$ as $g$ increases in order to minimize the energy,  and hence  favor pair formation, such that it will be of the paired form \eqref{Pairing} in its general bosonic version ($a^\dag_k a^\dag_{\bar k}\rightarrow (b^\dag_k)^2$).

On the other hand, since $[A,A^\dag]-1=2\sum_k\sigma_k^2 b^\dag_k b_k$, 
 for sp energies  $\varepsilon_k=\varepsilon\sigma_k^2$ and a fixed number of pairs $m=N/2\geq 2$,  
$H_B$ becomes proportional to the operator $-\hat  M_A$, Eq.\ \eqref{ncobd},   at 
\begin{equation}
g = g_c =\frac{\varepsilon}{m-1}\,.
\label{gc}
\end{equation}
Hence, at this value and for previous choice of sp energies, $H_B$ has a  pair condensate $\propto (A^\dag)^{m}|0\rangle$ as {\it exact nondegenerate} GS if $\varepsilon>0$, with energy $E^m_A=-mg_c=-\frac{m}{m-1} \varepsilon$. 

\begin{figure}[t]
\includegraphics[width=.95\linewidth]{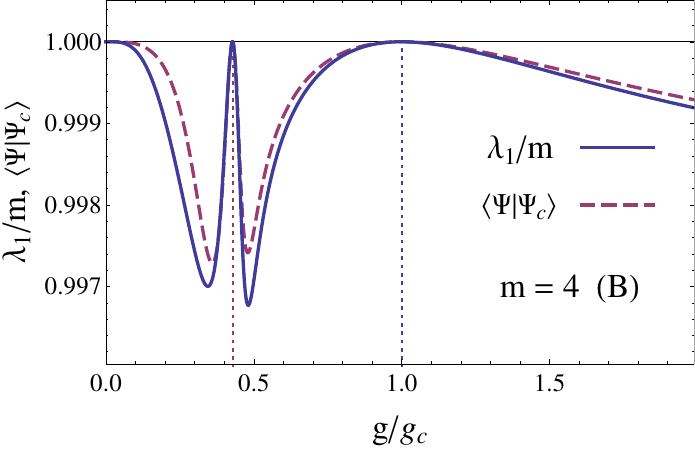}
\vspace*{-0.25cm}

\caption{
The largest eigenvalue $\lambda_1=\frac{1}{2}\lambda_{\rm max}$ of the effective density  $\frac{1}{2}\tilde{\bm\rho}_m^{(2)}$, Eq.\  \eqref{Lam2},  scaled by the number of pairs $m$ (blue solid line), in the exact GS $|\Psi\rangle$ of the bosonic  Hamiltonian \eqref{HBg}, as a function of the scaled coupling strength $g/g_c$, for $N=2m=8$ bosons. The dashed line depicts the overlap $\langle\Psi|\Psi_c\rangle$ between the exact GS and the pair  condensate $|\Psi_c\rangle\propto (\tilde A^\dag)^m|0\rangle$, with $\tilde A^\dag=\sum_{k,k'}\tilde A_{kk'}b^\dag_k b^\dag_{k'}$ and $\tilde{\bm A}$ the eigenvector associated to $\lambda_1$. The vertical dotted lines indicate the values of $g/g_c$ where the GS is exactly a pair condensate ($\lambda_1/m=\langle\Psi|\Psi_c\rangle=1$).}
\label{f1}
\vspace*{-.5cm}
\end{figure} 

Fig.\ \ref{f1} shows, as a function of $g/g_c$, the largest eigenvalue $\lambda_1$ of $\frac{1}{2}\tilde {\bm \rho}^{(2)}_m$, Eq.\ \eqref{Lam2}, scaled to $m$, in the GS of such $H_B$, together with the overlap $\langle \Psi|\Psi_c\rangle$ between the exact GS $|\Psi\rangle$ of $H_B$ and the condensate $|\Psi_c\rangle\propto (\tilde A^\dag)^m|0\rangle$, with $\tilde A^\dag$ obtained from the associated eigenvector of $\tilde{\bm\rho}_m^{(2)}$. 
We have considered $N=8$ bosons ($m=4$ pairs) in $n=N$ equally spaced sp levels $\varepsilon_k=\varepsilon k$,  with  $\sigma_k\propto \sqrt{k}$. 

As expected, it is first verified that $\lambda_1=m$ at $g=g_c$, where $\langle\Psi|\Psi_c\rangle=1$ and $\tilde A^\dag=A^\dag$. Thus, the exact GS of $H_B$ becomes exactly $\propto (A^\dag)^m|0\rangle$  at this {\it finite} value of $g$. Besides, the maximum value $\lambda_1=m$ is also reached at $g=0$ (no coupling), where all particles fall to the lowest level $\varepsilon_1$ and hence $\tilde A^\dag=(b^\dag_1)^2$: the GS becomes a standard condensate $\propto(b_1^\dag)^{2m}|0\rangle$ with energy $2m\varepsilon_1$. Since it is a  particular case of pair condensate, it is also  detected through the largest eigenvalue $\lambda_1$ of $\frac{1}{2}\tilde{\bm\rho}^{(2)}_m$. 

Remarkably, there is as well an intermediate {\it third point} where $\lambda_1=m$, which occurs here exactly at $g'_c=\frac{3}{7}g_c$. At this point the GS is again an {\it exact} pair condensate, as verified by the overlap 
$\langle \Psi|\Psi_c\rangle=1$. However, it is not generated by $A^\dag$, as here $\tilde A^\dag\propto \bar A^\dag$, 
with $\bar A^\dag\propto \sum_k \sigma_k^{-1}(b^\dag_k)^2$ the adjoint of the dual operator $\bar A$ of Eq.\ \eqref{Abar}.  
In order to understand this third point, we recall Eq.\ \eqref{Hbar}, which shows that the $A^\dag$ condensate can also emerge as a zero energy GS of a  Hamiltonian constructed with this partner operator $\bar{A}^\dag$. Then, replacing $\bar A^\dag, \bar A\rightarrow A^\dag,A$  in \eqref{Hbar}, it is seen that the Hamiltonian \eqref{HBg} will  exhibit a second nontrivial pair condensate GS $\propto (\bar A^\dag)^m|0\rangle$ with energy $E^{m}_{\bar A}=0$, at 
\begin{equation}
g'_c=\frac{m-1}{n/2+m-1}g_c\,,
\label{gcp}
\end{equation}
since at this value of $g$ it becomes proportional to \eqref{Hbar} with previous replacement. Eq.\ \eqref{gcp} holds for {\it any} choice of the $\sigma_k$. 

It is also observed in Fig.\ \ref{f1} that the exact GS remains quite close to a condensate for all $g$ values, since     $\langle\Psi|\Psi_c\rangle$ stays  above $\approx 0.9966$ in the whole interval considered. Moreover, this  overlap lies in this case very close to $\lambda_1/m$ for all $g$, exhibiting the same behavior, with minima in the vicinity of $g=g'_c$. Since $\lambda_1/m=1-D_2(|\Psi\rangle)/m$, 
with $D_2$ the proximity measure \eqref{DE}, we see that in this case $D_2(|\Psi\rangle)/m\approx 1-|\langle\Psi|\Psi_c\rangle|$, both vanishing exactly just at the points of exact pair condensation. 

\begin{figure}[t]
\vspace*{-.25cm}
\includegraphics[width=.8\linewidth]{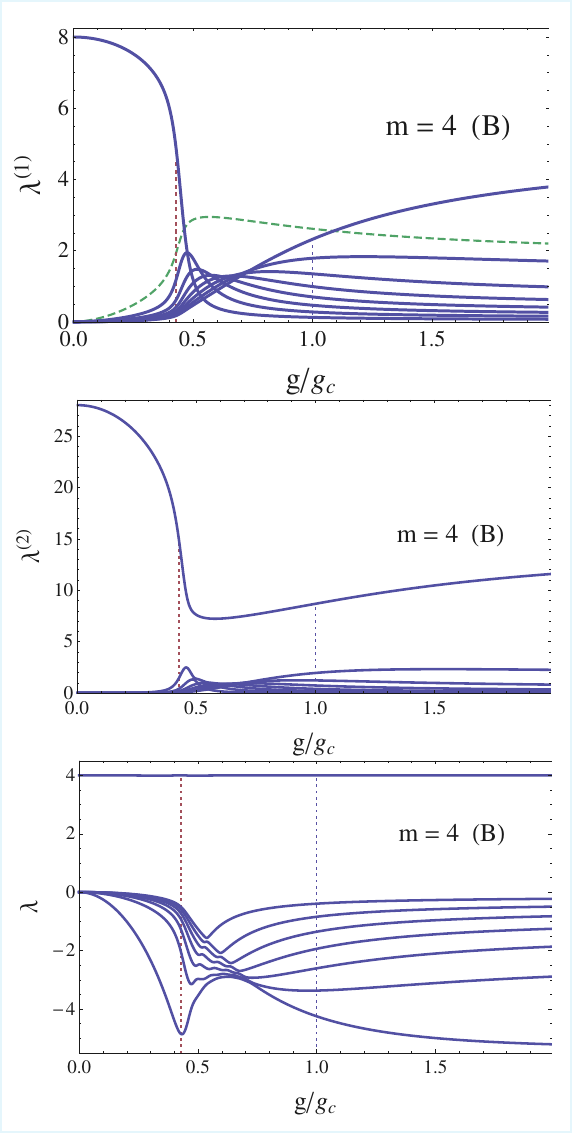}
\vspace*{-0.25cm}
\caption{
The eigenvalues of the one-body (top) and two-body (center) density matrices, and those of the effective density $\frac{1}{2}\tilde{\bm\rho}_m^{(2)}$ (bottom), Eq.\  \eqref{Lam2}, as a function of $g/g_c$ in the GS of the bosonic Hamiltonian \eqref{HBg}, for the same case of Fig.\ \ref{f1}.  Vertical dotted lines indicate the values of $g/g_c$ where exact GS pair condensation takes place. In the top panel the associated one-body entropy $S(\bm\rho^{(1)}_n)$ (dashed line) is also depicted.}
\label{f2}
\vspace*{-.5cm}
\end{figure}

Further understanding of the GS behavior can be obtained from 
the eigenvalues of the one- and two-body DMs $\bm\rho^{(1)}$ and $\bm\rho^{(2)}$, Eqs.\ \eqref{C10}--\eqref{C20}, and those of $\frac{1}{2}\tilde{\bm\rho}_m^{(2)}$, depicted in Fig.\ \ref{f2}. 

In the top panel of Fig.\ \ref{f2} it is first seen that 
the average occupations of the natural orbitals, given by the eigenvalues $\lambda_k^{(1)}=\langle b^\dag_k b_k\rangle$ of $\bm\rho^{(1)}$, undergo  an inversion as the coupling strength $g$ increases: Starting from a standard condensate at $g=0$, where all bosons are in the lowest sp level  ($\lambda_k^{(1)}=2m\delta_{k1}$), the average occupation ordering remains opposite to the sp level ordering ($\lambda^{(1)}_k>\lambda^{(1)}_{k'}$ if $\varepsilon_k<\varepsilon_{k'}$) for $g/g_c\alt 1/2$, i.e., in the weak coupling regime. Accordingly, it is in this sector where we find the $\bar A$ condensate as exact GS, since in this condensate occupations are approximately proportional to $\sigma_k^{-2}\propto\varepsilon_k^{-1}$. 
Nevertheless, as $g$ increases the attractive coupling $-gA^\dag A$, which favors the inverse occupation ordering,  
prevails, and the complete population inversion takes  place for $g/g_c\agt 0.75$. Accordingly, the $A^\dag$ condensate is located in this last sector, as it implies  the opposite ordering  ($\lambda^{(1)}_k>\lambda^{(1)}_{k'}$ if $\varepsilon_k>\varepsilon_{k'}$). 

We also depict in the top panel (dashed line) the associated one-body entanglement entropy \cite{CR.24,GDR.20}   
$S(\bm\rho^{(1)}_n)$, where $\bm\rho^{(1)}_n=\bm\rho^{(1)}/N$ is the normalized one-body DM and $S(\rho)=-{\rm Tr}\,\rho\log_2\rho$ the von Neumann entropy. It is here maximum 
in the transition region between both occupation orderings (i.e.\ where the eigenvalues $\lambda_k^{(1)}$ are most uniform)  and not at the points of exact pair condensation,  nor in the limit of strong couplings $g\gg g_c$ (as occurs for a plain uniform $A^\dag$ \cite{CR.24,DGR.18}). 

On the other hand, the eigenvalues of the two-body DM $\bm\rho^{(2)}$,  shown in the central panel,  exhibit a dominant largest eigenvalue $\lambda_1^{(2)}$ characteristic of pairing-type correlations \cite{CR.24}: While its maximum is reached at  the $g=0$ standard condensate limit ($\lambda^{(2)}_k=\frac{1}{2}\langle b^{\dag\,2}_{k}b_k^2\rangle=\delta_{k1} m(2m-1)$),   it remains large and  well detached from the remaining  eigenvalues for all $g>0$, 
becoming minimum in the previous transition region. Whereas the presence of a dominant  eigenvalue in $\bm\rho^{(2)}$ certainly indicates approximate 
condensate-like behavior of the GS, no special signature is exhibited  by this eigenvalue (nor by the others)  at the points (vertical dotted lines) where the GS is an exact condensate. 
Hence, it cannot directly detect the point of exact GS pair  condensation.

The eigenvalues $\lambda$ of the modified DM \eqref{Lam2} are shown in the bottom panel. It is seen that its largest eigenvalue, which is that detecting exact pair condensation,  is here the only positive one 
(and almost constant with $g$ when shown in this larger scale), so that it is well separated from the rest.  
We remark that in the case of $\bm\rho^{(2)}$ (and $\tilde{\bm\rho}^{(2)}_m$) we have just depicted   the eigenvalues of the ``collective'' block of these matrices (containing the elements $\frac{1}{2}\langle b^{\dag\,2}_k b_{l}^2\rangle$ in the natural basis), which is that leading to the largest eigenvalue. Remaining blocks of $\bm\rho^{(2)}$ are here diagonal, as 
$\langle b^\dag_k b^\dag_l b_{l'} b_{k'}\rangle=\delta_{kk'}\delta_{ll'}\langle b^\dag_k b^\dag_l b_l b_k\rangle$ for $k<l$, $k'<l'$ in the present GS, and are 
 irrelevant for determining its  largest eigenvalue. 

Regarding the relation with previous panel, we note that the largest eigenvalue $\lambda_1^{(2)}$ of $\bm\rho^{(2)}$ in a pair boson condensate is always $\geq m$ in boson systems (actually $\lambda_1^{(2)}\geq m(1+2\frac{m-1}{n})$ \cite{CR.24}, this minimum value reached in the uniform case $A^\dag=A^\dag_0$, Eq.\ \eqref{eigenA0}), and maximum in a standard condensate.  Hence, proximity to a  pair condensate is associated to a large maximum eigenvalue  of $\bm\rho^{(2)}$ in the boson case, as verified in the central and bottom panels,   and hence to off-diagonal long range order (ODLRO) \cite{Yang.62}  if signaled by such large eigenvalue. The nature of the associated pair can always be derived from the associated eigenvector, which determines $A^\dag$. 

\subsection{Fermionic system}
In the fermionic case we consider an  analogous pairing Hamiltonian
\begin{equation}
H_F=\tfrac{1}{2}\sum_k\varepsilon_k (a^\dag_ka_k+a^\dag_{\bar k}a_{\bar k})-gA^\dag A\,,\label{HFg}
\end{equation}
where $A^\dag=\sum_k \sigma_k a^\dag_k a^\dag_{\bar k}$,  $\sum_k \sigma_k^2=1$ and $k=1,\ldots,n/2$.  For  attractive coupling $g>0$, its  GS will again be of the general paired form \eqref{Pairing}, with positive coefficients $\Gamma_{m_1
\ldots m_d}$. 

For $\varepsilon_k=-\varepsilon \sigma_k^2$ and fixed pair number $m=N/2\geq 2$, $H_F$ will become  proportional to $-\hat M_A/(m-1)$, with $\hat M_A$ the (now fermionic)  operator  \eqref{ncobd}, at the {\it same} value \eqref{gc} of the coupling $g$. At this  point its GS is then an exact pair condensate $\propto (A^\dag)^m|0\rangle$ for each 
value of $m$ (and $\varepsilon>0$), again with energy $E^m_A=-\frac{m}{m-1}\varepsilon$. 

\begin{figure}[t]
\includegraphics[width=.95\linewidth]{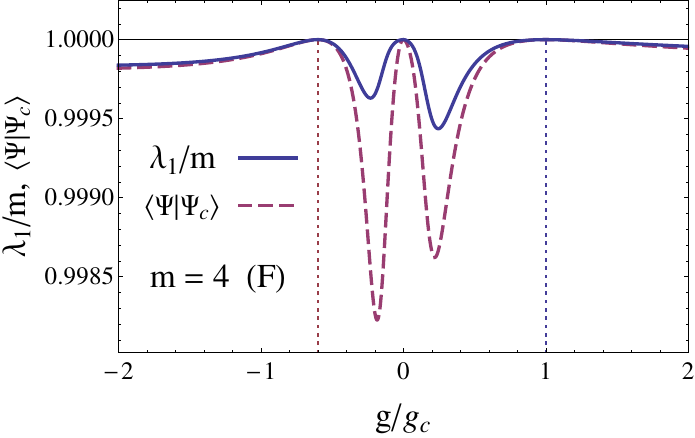}
\vspace*{-0.25cm}

\caption{ 
Same details as Fig.\ \ref{f1} in the  fermionic case, for Hamiltonian \eqref{HFg} and $N=2m=8$ fermions. Here $g/g_c<0$ indicates $g>0$ but $\varepsilon<0$ (opposite sp spectrum) in \eqref{HFg}.} 
\label{f3}
\vspace*{-.5cm}
\end{figure} 
   
We also notice that in the fermionic case the second nontrivial condensate $\propto (\bar A^\dag)^m|0\rangle$ is eigenstate  of $H_F$ for an  opposite sp spectrum $\varepsilon_k=+\varepsilon\sigma_k^2$, at  
\begin{equation}
g'_c=\frac{m-1}{n/2-(m-1)}g_c\,,
\label{gpcf}
\end{equation}
 with energy $E^m_{\bar A}=0$, since for this value and spectrum $H_F$  becomes proportional to \eqref{Hbar}. This condensate will be GS if $\varepsilon>0$. Here $n$ is the total number of sp states. 

The corresponding GS results for the highest eigenvalue $\lambda_1$ of $\frac{1}{2}\tilde{\bm \rho}^{(2)}_m$ and the ensuing overlap $\langle\Psi|\Psi_c\rangle$ are shown in Fig.\ \ref{f3} for a system of $N=8$ fermions ($m=4$ pairs) in  $n=16$ sp states, again with an equally spaced sp spectrum $\varepsilon_k=-\varepsilon k$ and $\sigma_k\propto \sqrt{k}$, $k=1,\ldots n/2$. In order to also expose the second condensate in the same figure, we have included negative values of $g/g_c$, which mean $g>0$ but $\varepsilon<0$ in \eqref{HFg} (i.e.\ $\varepsilon_k>0$) such that  it arises at $g/g_c=-|g'_c/g_c|$, i.e.\ $-3/5$ in the case considered. 

It is verified in Fig.\ \ref{f3} that $\lambda_1$  again  reaches its maximum $m$ at $g=g_c$ ($A^\dag$ condensate, $\tilde A^\dag=A^\dag$),  at $g=0$, where the GS is a SD and hence it can be written as $\propto ({\tilde A}^\dag)^m|0\rangle$ with ${\tilde A}^\dag=\sum_{k=1}^{m}c^\dag_kc^\dag_{\bar k}$ (sum over the $m$ lowest, sp levels), and at $g/g_c=-\frac{3}{5}$ as previously stated, where $\tilde A^\dag\propto\bar A^\dag$ and the GS is $\propto ({\bar A}^\dag)^m|0\rangle$. The behavior of the overlap $\langle\Psi|\Psi_c\rangle$ follows again that of $\lambda_1/m$, 
becoming of course $1$ when $\lambda_1=m$, but is now lower,  especially at the minima of $\lambda_1$. Nonetheless,  its value remains  quite  high for all values of $g$, reflecting  the proximity of the exact GS to a condensate for any $g$. 

\begin{figure}[t]
\vspace*{-0.25cm}

\includegraphics[width=.8\linewidth]{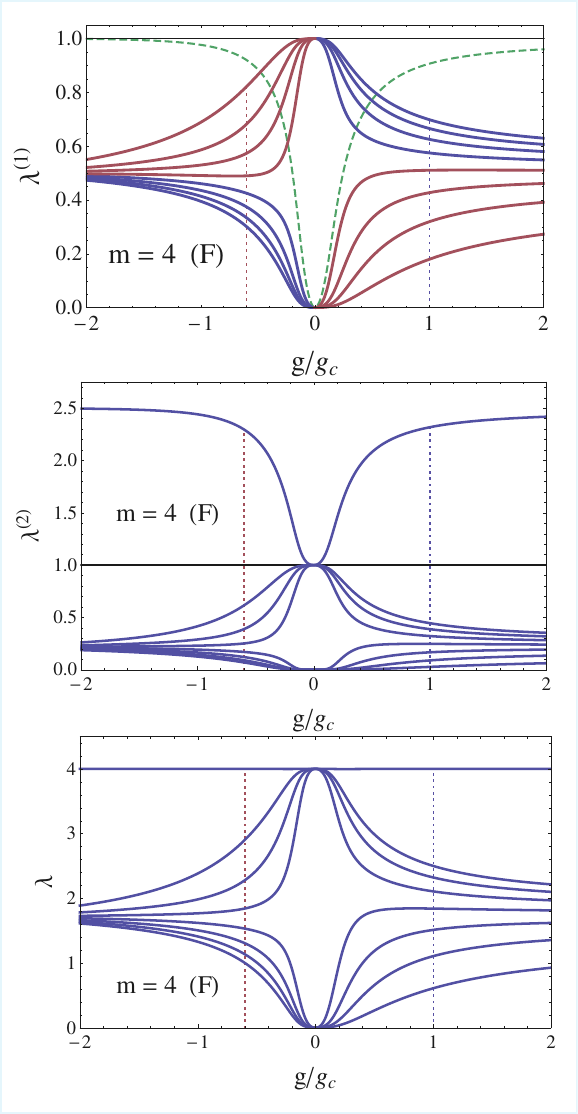}
\vspace*{-0.5cm}
\caption{
Same details as Fig.\ \ref{f2} in the fermionic case, for the GS of Hamiltonian \eqref{HFg} in the same case of Fig.\ \ref{f3}. In the top panel the blue (red) lines depict the average occupation of the lowest (highest) sp levels for $g/g_c>0$. Their ordering is reversed for $g/g_c<0$, where the sp levels change sign. The dashed line shows the associated one-body entropy.}
\label{f4}
\vspace*{-.5cm}
\end{figure}

Further understanding of the fermionic GS results can be obtained from Fig.\ \ref{f4}. 
The top panel depicts the  eigenvalues (here two-fold degenerate due to the $k-\bar k$ degeneracy) of the fermionic one-body DM $\bm\rho^{(1)}$ in the GS of Hamiltonian \eqref{HFg}. 
Due to the minus sign in the sp spectrum for $g/g_c>0$ in \eqref{HFg},  the average occupation ordering of the natural orbitals now follows that favored by $A^\dag$, i.e., by the attractive interaction $-gA^\dag A$,  for all $g>0$: $\lambda^{(1)}_k> \lambda^{(1)}_{k'}$ if $|\varepsilon_k|> |\varepsilon_{k'}|$, i.e.\ $\sigma_k> \sigma_{k'}$, so that there is no occupation inversion as $g/g_c$ increases from $0$, as seen in the top panel.  Therefore, just the $A^\dag$ condensate GS arises here for $g>0$.  

The partner GS condensate $\propto({\bar A^\dag})^m|0\rangle$ 
emerges instead for  negative values of $g/g_c$, where the spectrum becomes inverted due to the sign change of $\varepsilon$ 
and hence the occupation ordering for sufficiently weak  $g/g_c<0$ is that favored by 
 $\bar A^\dag$ ($\lambda^{(1)}_k< \lambda^{(1)}_{k'}$ if $|\varepsilon_k|> |\varepsilon_{k'}|$) instead of $A^\dag$.   Occupation ordering inversion will take place for   higher negative values of $g/g_c$ (here exactly at $g/g_c=-3$, where all sp occupations merge: $\lambda_k^{(1)}=1/2$ $\forall\,k$). 
 
It is also seen that all levels become occupied on average as $|g/g_c|$ increases, reflecting the departure of the GS from a SD and hence the increase of the one-body entanglement entropy \cite{GDR.20,CR.24},  depicted as well in the top panel. We plot here $\Delta S_n=S(\bm 
\rho^{(1)}_n)-\log_2 N$, such that $\Delta S_n=0$  for a SD and hence at $g/g_c=0$. It becomes maximum for a uniform spectrum, 
i.e.\  $\lambda^{(1)}_k=\frac{1}{2}$ $\forall\,k$ in the present half-filled case, where $\Delta S_n=1$ (this value is here reached at the previous inversion point).   

The spectrum of $\bm\rho^{(2)}$, depicted  in the central panel of Fig.\ \ref{f4}, shows the emergence of a large dominant eigenvalue ($\lambda^{(2)}_1>1$) as $|g/g_c|$ increases from $0$, reflecting the onset of pairing correlations \cite{CR.24}, though no special feature is exhibited at the points of exact GS pair condensation. On the other hand, those of the effective DM $\frac{1}{2}\tilde{\bm\rho}^{(2)}_m$ (bottom panel) are now all positive, since in the fermionic case it is positive semidefinite, as seen from Eq.\ \eqref{Lam2}.  Nonetheless, its largest eigenvalue $\lambda_1$ lies again well detached from the rest if $|g/g_c|$ is not small, and is  almost constant at this larger scale. The main difference with the bosonic case is that it becomes degenerate in the $g\rightarrow 0$ limit, where it merges with all remaining nonzero eigenvalues, acquiring the same degeneracy as the largest eigenvalue of $\bm\rho^{(2)}$  ($\binom{N}{2}$ for a $N$-particle SD; as in the bosonic case, we have just depicted in Fig.\ \ref{f4} those of the ``collective'' block of $\bm\rho^{(2)}$ and $\frac{1}{2}\tilde{\bm\rho}_m^{(2)}$, containing the contractions 
$\langle c^\dag_k c^\dag_{\bar k}c_{\bar k'}c_{k'}\rangle$ and hence the dominant largest eigenvalue $\lambda_1^{(2)}$ and $\lambda_1$ of these matrices). Thus, when $\lambda_1=m$, ``true'' fermionic pair condensates can be easily distinguished from SDs  just by considering its degeneracy, as previously discussed. 

\begin{figure}[t]
\includegraphics[width=.9\linewidth]{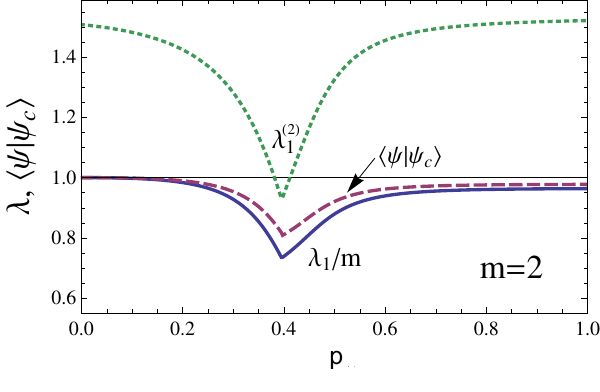}
\vspace*{-0.5cm}

\caption{Same details of Fig.\ \ref{f3} for the fermionic Hamiltonian \eqref{HFgp}   (see text) and $N=4$ fermions.   Its  GS  is an exact  coboson condensate just at $p=0$, evolving to a distinct paired GS  for increasing $p\rightarrow 1$. While the largest eigenvalue $\lambda_{1}^{(2)}$ of the two-body DM $\bm\rho^{(2)}$ (dotted line) is large ($>1$) in both limits, reflecting pairing, that of \eqref{Lam2} stays   close but below $m$ at the right limit, indicating  deviation of the GS  from an exact  pair condensate, as verified by the overlap $\langle\Psi|\Psi_c\rangle<1$. In the transition region all three quantities depicted exhibit a pronounced minimum, reflecting a strong deviation from a pair condensate.} 
\label{f5}
\vspace*{-.5cm}
\end{figure}

In fermionic pair condensates,  the largest eigenvalue 
of $\bm\rho^{(2)}$  
 satisfies $1\leq \lambda_1^{(2)}\leq m(1-2\frac{m-1}{n})$ \cite{CR.24}, 
 the upper bound  reached at the uniform condensate $A^\dag=A^\dag_0$  \cite{Yang.62} (Eq.\ \eqref{eigenA0}).  Hence, while proximity to a pair condensate implies in general $\lambda_1^{(2)}\geq 1$,  in the fermionic case this maximum eigenvalue is not necessarily of order $m$ (it becomes in fact $1$ in the limit of a SD, as seen in the central panel) such that ODLRO is not ensured by this proximity. 

For completeness, we finally  show in Fig.\ \ref{f5} results for the  GS of a Hamiltonian 
\begin{equation}
H'_F=(1-p)H_{F_1}+pH_{F_2},
\label{HFgp}
\end{equation}
where both $H_{F_1}$ and $H_{F_2}$ are of the form  \eqref{HFg} but in different sp basis, with $g=g_c$ in $H_{F1}$ and $g\neq g_c$ in $H_{F2}$. Thus, its GS becomes an exact pair condensate for $p\rightarrow 0$, where both $\lambda_1/m$ and the overlap $\langle\Psi|\Psi_c\rangle$ approach $1$, but not for $p\rightarrow 1$, where these quantities become  just close to $1$. For intermediate values of $p$, we see that both $\lambda_1/m$ and the overlap acquire values well below $1$, reflecting no proximity to a pair condensate, and also  no pairing, as the largest eigenvalue of $\bm\rho^{(2)}$, well above $1$ for both $p\rightarrow 0$ and $p\rightarrow 1$,  also becomes here less than $1$.  A transition between distinct GS regimes is exhibited at $p\approx 0.4$ in both $\lambda_1$ and $\lambda_1^{(2)}$, as well as the overlap, through a slope discontinuity.

Thus, 
through the largest eigenvalue of $\tilde{\bm \rho}^{(2)}_m$ and the corresponding eigenvector, 
one can detect the proximity of the GS to an actual pair condensate as well as the nature of the ``closest''  pair 
condensate, allowing one to identify distinct GS regimes. This could be applied, for instance, to problems such as the BCS--BEC crossover \cite{MP.14,PL.07}, at least within simple pair condensate-based descriptions. We also remark that 
 if $\bm\rho^{(1)}$ and $\bm\rho^{(2)}$  come from a state $f(A^\dag)|0\rangle$ with no fixed particle number but fixed even number parity, like e.g.\ quasiparticle vacua $\propto e^{\alpha A^\dag}|0\rangle$, the  present scheme  can also exactly detect them with Eqs.\ \eqref{Condition2f}--\eqref{rt},  and determine the pertinent $A^\dag$ through the corresponding  eigenvector. The largest eigenvalue of the r.h.s. in \eqref{Condition2f} could also be employed to estimate the proximity to any such state and its eigenvector for obtaining  an ``optimum'' $A^\dag$.  Finally, we recall that the method does not rely on bosonic properties of $A^\dag$. Once obtained from such eigenvector, its  bosonic and correlation properties can be evaluated through suitable measures (like the $\chi_N$ ratio  \cite{CKL.05,COW.10} and the pair entanglement entropy).

\section{Conclusions\label{IV}}
We have presented a novel characterization of exact pair condensates in both boson and fermion systems, through the identification of  the associated set of  conserved one-body operators, i.e., operators which have such states as exact eigenstate. The dimension of this subspace of operators, typically  very low for  random states, has unique maximal properties for these pair condensates when considering correlated states with full rank one-body densities (without ``frozen'' levels in the fermionic case), being independent of the  number $m$ of pairs.  Through this set we were also able to construct the most general two-body Hamiltonian having such condensates as eigenstate, including a subset of Hamiltonians which have them as ground state. They include as special cases known 
pairing-like Hamiltonians with special couplings, but is not limited to them. 

 By means of the present scheme we could also identify a simple necessary and sufficient  condition for detecting an exact pair condensate from the  knowledge of its one- and two-body DMs, which also yields the relevant pair operator $A^\dag$, thus enabling the exact reconstruction of the state.  This condition also provides a simple measure of the proximity of a given state to a pair condensate,  
together with a ``nearest'' pair creation operator  and condensate, which minimize a related average energy.  
As shown in the examples, the formalism is  useful for rapidly detecting when the  GS of a given Hamiltonian becomes an exact pair condensate, or for determining its proximity to a pair condensate. Extension of the present scheme to more complex states is under investigation. 

\begin{acknowledgments} 
Authors acknowledge support from CONICET (FP)  and CIC (RR) of Argentina.  Work supported by CONICET PIP Grant No. 11220200101877CO.
\end{acknowledgments}

\appendix
\section{Uniform case and proof of Proposition 1$\!\!$\label{ApA}}

Let us first consider the uniform case, where  all $\sigma_{k}$ in Eq.\ \eqref{Schmidt} are equal. We then obtain a perfect ladder operator 
\begin{subequations}
\begin{eqnarray}
A_{0}^{\dagger} &=& \sqrt{\tfrac{2}{n}}  \sum_{k=1}^{n/2}a_{k}^{\dagger}a_{\bar{k}}^{\dagger},\\
A_{0}^{\dagger}&=&\tfrac{1}{\sqrt{2 n}} \sum_{k=1}^{n}b_{k}^{\dagger2},
\end{eqnarray}
\label{A0}
\end{subequations}
in the fermionic and bosonic case respectively, satisfying
\begin{equation}
[A_{0},A_{0}^{\dagger}]=1\mp\tfrac{2}{n}\hat{N}\,,
\label{commutationA0}
\end{equation}
and $[\hat N,A_0^\dag]=2A_0^\dag$.  
Eq. \eqref{commutationA0} implies
\begin{equation}
[A_{0},(A_{0}^{\dagger})^{m}]=m(A_{0}^{\dagger})^{m-1}[1\mp\tfrac{2(\hat{N}+m-1)}{n}]\,.
\label{commutationA0m}
\end{equation}
Hence, the states   $|m_0\rangle_2=\tfrac{1}{\sqrt{\mathcal{N}_m}}(A_0^{\dagger})^{m}|0\rangle$ satisfy \cite{CR.24}
\begin{subequations}
\begin{eqnarray}
A_0^{\dag}|m_0-1\rangle_2&=& \sqrt{m\left(1\mp\tfrac{2(m-1)}{n}\right)}|m_0\rangle_2\,,
\label{ladder}\\
A_0^{\dagger}A_0|m_0\rangle_2&=&m\left(1\mp\tfrac{2(m-1)}{n}\right)
|m_0\rangle_2\,,
\label{eigenA0}
\end{eqnarray}
\end{subequations}
being the non-degenerate GS of $-A_{0}^{\dagger}A_{0}$ within each $N=2m$ subspace. Eq.\  \eqref{eigenA0} is a particular case of Eq.\ \eqref{ncob}, and can be directly obtained from \eqref{ncobd} using \eqref{commutationA0}.  

These relations can also be obtained from the well known seniority scheme for fermion pairing (see e.g.\ \cite{RS.80,LWS.80}).  In fact, for fermions (bosons), the operators  $S_+=\sqrt{\frac{n}{2}}A_0^\dag$, $S_-=S_+^\dag$ and $S_z=\frac{1}{2}\hat N\mp\frac{n}{4}$ satisfy an  $SU(2)$ ($SU(1,1)$) algebra, i.e.\ $[S_z,S_{\pm}]=\pm S_{\pm}$, $[S_+,S_-]=2S_z$ ($-2S_z$) \cite{RS.80,LWS.80,DES.01}. 
Thus, for fermions 
$|m_0\rangle_2$ corresponds to 
$|S,M\rangle$, with 
$S=\frac{n}{4}$, $M=m-\frac{n}{4}$.  

In the general case, a  pair creation operator \eqref{Schmidt} with maximum rank 
can be obtained from \eqref{A0} through the transformation
\begin{equation}
A^{\dagger}=e^{-h}A_{0}^{\dagger}e^{h}\,,\label{A7}
\end{equation}
where $h$ is the hermitian one-body operator 
\begin{subequations}
\begin{eqnarray}
h &=&-\tfrac{1}{2}\sum_{k=1}^{n/2}\ln(\sigma_{k})(a_{k}^{\dagger}a_{k}+a_{\bar{k}}^{\dagger}a_{\bar{k}})\,,\\
h &=&-\tfrac{1}{2}\sum_{k=1}^{n}\ln(\sigma_{k})b_{k}^{\dagger}b_{k}\,,
\end{eqnarray}
\end{subequations}
for fermions and bosons respectively, such that 
\begin{subequations}
\begin{eqnarray}
e^{-h} a_{k,\bar{k}}^{\dagger}e^{h} &=& \sqrt{\sigma_{k}}\,a_{k,\bar{k}}^{\dagger}\,,\\
e^{-h}b_{k}^{\dagger}e^{h} &=& \sqrt{\sigma_{k}}\,b_{k}^{\dagger}\,,
\end{eqnarray}
\end{subequations}
and $e^{-h}a_{k,\bar k}e^h=a_{k,\bar k}/\sqrt{\sigma_k}$, 
$e^{-h}b_{k}e^h=b_k/\sqrt{\sigma_k}$. 
Since $e^h|0\rangle=|0\rangle$,  \eqref{A7} implies the following relation  between the general  $|m\rangle_2$ and the uniform $|m_0\rangle_2$ pair condensates: 
\begin{equation}
|m\rangle_2\propto e^{-h} |m_0\rangle_2\,.
\label{trans}
\end{equation}

In particular, Eq.\ \eqref{trans} entails that  a conserved quantity $Q$ associated to $|m\rangle$ is related to  a
conserved quantity $Q_{0}$ associated to $|m_{0}\rangle$ through 
\begin{equation}
Q= e^{-h}Q_{0}e^{h},
\label{transQ}
\end{equation}
such that $Q|m\rangle_2=\lambda |m\rangle_2$ iff $Q_0|m_0\rangle=\lambda|m_0\rangle$.  This relation can be verified, for instance, in Eqs. \eqref{Qconservedf}-\eqref{Qconservedb}. 

Similarly, if $\bar Q$ is a conserved quantity associated to $|\bar m\rangle \propto ({\bar A}^\dag)^m |0\rangle$, with $\bar A$ the dual operator \eqref{Abar}, then  
\begin{equation}
\bar Q\propto e^{h}Q_{0}e^{-h}\,,
\label{transQbar}
\end{equation}
since 
\begin{equation}
\bar A^\dag=e^h A^\dag_0 e^{-h}\label{Abart}
    \end{equation}
such that $\bar{A}=e^{-h}A_0 e^{h}$. 
This ensures that $[\bar A,A^\dag]=e^{-h}[A_0,A_0^\dag]e^{h}=[A_0,A_0^\dag]$, Eq.\ \eqref{comAbar}.

{\it Proof of Proposition 1.} Eq.\ \eqref{trans} 
also allows us to easily prove the fermionic identity  of Eq.\ \eqref{Statebis}:  Since, as previously stated, for fermions $|m_0\rangle_2$ is equivalent to a state $|S,M\rangle\propto S_+^{S+M}|S,-S\rangle$, with  $S=\frac{n}{4}$, $M=m-\frac{n}{4}$, $S_+\propto A_0^\dag$ and $|S,-S\rangle=|0\rangle$, it can also be 
expressed as $|S,M\rangle\propto S_-^{S-M}|S,S\rangle$, i.e.\ 
\begin{equation}
|m_{0}\rangle_2=\tfrac{1}{\sqrt{\bar{\mathcal{N}}}_m}(A_{0})^{\tfrac{n}{2}-m}|\bar{0}\rangle\,,\label{A12}
\end{equation}
as $|S,S\rangle$ corresponds to  the ``maximally occupied'' state $|\bar{0}\rangle$ ($m=n/2$).  
Then, for a general $A^\dag$, Eq.\ \eqref{Statebis} follows from \eqref{A7} and \eqref{trans}, as   $e^{-h}A_0 e^h=\bar A$ (Eq.\ \eqref{Abart})  while $e^{-h}|\bar{0}\rangle\propto |\bar{0}\rangle$. \qed

\section{proof of Theorem 1 and the algebra of conserved one-body operators\label{ApB}}

We consider conserved quantities of the form
\begin{equation}
Q=\sum_{i,j} h_{ij} c_i^\dag c_j.
\end{equation}
Since the number operator is a trivial conserved quantity of this kind,  satisfying $\hat{N} |m\rangle_2 = 2m |m\rangle_2$, we have  $Q|m\rangle_2 = \lambda_m |m\rangle_2$ iff 
\begin{equation}
\tilde{Q} |m\rangle_2 = 0,
\label{conservedQm}
\end{equation}
where $\tilde{Q} = Q - \frac{\lambda_m}{2m} \hat{N} = \sum_{i,j} \tilde{h}_{ij} c_i^\dag c_j$ and 
 $\tilde{h}_{ij}=h_{ij}-\frac{\lambda_m}{2m} \delta_{ij}$. Since
\begin{equation}
[\tilde{Q},A^\dag] = \tfrac{1}{2} (\tilde{{\bf h}} {\bf A} \mp (\tilde{\bf h} {\bf A})^t) c_i^\dag c_j^\dag\,,
\end{equation}
is a two particle creation operator satisfying $[[\tilde{Q},A^\dag],A^\dag] = 0$, Eq. \eqref{conservedQm} leads to
\begin{equation}
[\tilde{Q},A^\dag] (A^\dag)^{m-1} |0\rangle = 0\,,
\end{equation}
implying that $[\tilde{Q},A^\dag]$ is a conserved quantity of $|m-1\rangle_2$. Thus, due to Proposition 2, for $m \leq n/2-1$ in fermions and for all $m$ in bosons, we arrive at \begin{equation}[\tilde{Q},A^\dag]=0\label{CQA}\end{equation} implying 
\begin{equation}
\tilde{{\bf h}} {\bf A} = \pm (\tilde{\bf h} {\bf A})^t.
\label{eqh}
\end{equation}
Since $\bf A$ is non singular, we can define ${\bf M} = {\bf A}^{-1} \tilde{{\bf h}}$ and then, Eq. \eqref{eqh} implies that ${\bf M} = \pm {\bf M}^t$. Finally, we arrive at $\tilde{{\bf h}} = \bf{A} \bf{M}$ with $\bf{M}$ an arbitrary symmetric (skew-symmetric) matrix, implying $\tilde{Q} =- \tfrac{1}{2} \sum_{i,j} M_{ij} Q_{ij}$, where the $Q_{ij}$ are given by \eqref{Qconserved}. Therefore, they span  the whole space of conserved quantities of this type.

Furthermore, for $A^\dag=A^\dag_0$ and fixed  $N=2m$, Eq. \eqref{ncob} leads to \eqref{eigenA0} and it is well known that the unique eigenstate of $A_0^\dag A_0$ having $m\left[1\mp\tfrac{2(m-1)}{n}\right]$ as eigenvalue is $|m_0\rangle_2$ (for $N$ odd this is no longer an eigenvalue). Thus, for this case, we can claim that $H_{A_0}|\psi\rangle = \tfrac{1}{4} \sum_{i,j} (Q^0_{ij})^\dag Q^0_{ij} |\psi\rangle = 0$ implies $|\psi\rangle = |m_0\rangle_2$ (since $H_{A_0}=-A_0^\dag A_0$ plus constant terms for fixed $N$), and then $Q_{ij}^0 |\psi\rangle = 0 \ \forall \ i,j$ implies $|\psi\rangle=|m_0\rangle$. In the general case, $Q_{ij}|\psi\rangle=0 \ \forall \ i,j$ implies $Q_{ij}^0 e^h |\psi\rangle=0 \ \forall \ i,j$ and then $e^h |\psi\rangle \propto |m_0\rangle_2$ due to previous result. Hence, we finally obtain $|\psi\rangle \propto e^{-h} |m_0\rangle_2 = |m\rangle_2$. 
Therefore,  the conserved operators $Q_{ij}$ and the number operator $\hat N$ define the state univocally. \qed

As a check, it is straightforward to verify that the full set of conserved one-body  operators $Q_{ij}$  is closed under commutation: They satisfy
\begin{equation}
\!\!\![Q_{ij},Q_{kl}]=\pm(A_{ki}Q_{jl}\!+\!A_{lj}Q_{ik})\mp(A_{jk}Q_{il}\!+\!A_{il}Q_{jk})\,.\label{AlgC}    
\end{equation}

Moreover, the ``normal'' conserved operators \eqref{Qconservedf}--\eqref{Qconservedb} satisfy essentially $SU(2)$ commutation relations for each pair $k,l$ when adequately scaled: 
In  the fermionic case,  
\begin{equation}
\begin{split}
S_{kl}^+&=\tfrac{Q_{\bar k \bar l}}{\sqrt{\sigma_k\sigma_l}}=\sqrt{\tfrac{\sigma_k}{\sigma_l}} a_k^\dag a_{\bar l}+\sqrt{\tfrac{\sigma_l}{\sigma_k}} a_{l}^\dag a_{\bar k}\,,\\
S_{kl}^-&=\tfrac{Q_{kl}}{\sqrt{\sigma_k\sigma_l}}=\sqrt{\tfrac{\sigma_k}{\sigma_l}} a_{\bar k}^\dag a_l+\sqrt{\tfrac{\sigma_l}{\sigma_k}} a_{\bar l}^\dag a_k\,,\\
S_{kl}^z&=\tfrac{Q_{\bar k k}}{2\sigma_k}+\tfrac{Q_{\bar l l}}{2\sigma_l}=\tfrac{1}{2}
(a^\dag_k a_k-a^\dag_{\bar k}a_{\bar k}+a^\dag_l a_l-a^\dag_{\bar l}a_{\bar l}),
\label{B8}
\end{split}
\end{equation}
satisfy, for $k<l$,  
\begin{equation}[S_{kl}^+,S_{kl}^-]=2S_{kl}^z\,,\;\;[S^z_{kl},S^{\pm}_{kl}]=\pm S^{\pm}_{kl}\,.\label{SU2}\end{equation} 
The same relations are fulfilled, for $k<l$, by 
\begin{equation}
\begin{split}
\!\!\!S_{\bar k l}^+&=\tfrac{Q_{\bar k l}}{\sqrt{\sigma_k\sigma_l}}=\sqrt{\tfrac{\sigma_k}{\sigma_l}} a_k^\dag a_l-\sqrt{\tfrac{\sigma_l}{\sigma_k}} a_{\bar l}^\dag a_{\bar k}\,,\\
\!\!\!S_{\bar k l}^-&=\tfrac{Q_{
\bar l k}}{\sqrt{\sigma_k\sigma_l}}=\sqrt{\tfrac{\sigma_l}{\sigma_k}} a_l^\dag a_k-\sqrt{\tfrac{\sigma_k}{\sigma_l}} a_{\bar k}^\dag a_{\bar l}\,,\\
\!\!\!S_{\bar k l}^z&=\tfrac{Q_{\bar k k}}{2\sigma_k}-\tfrac{Q_{\bar l l}}{2\sigma_l}=\tfrac{1}{2}
(a^\dag_k a_k-a^\dag_l a_l+a^\dag_{\bar l}a_{\bar l}-a^\dag_{\bar k}a_{\bar k})\,.
\end{split}
\label{B10}
\end{equation}

 These operators are related to  those  of the uniform case $\sigma_k=\frac{1}{\sqrt{n}}\,\forall\,k$ by the similarity transformation 
\eqref{transQ}, then having the same eigenvalues as the standard angular momentum operators, even though $(S^-_{kl})^\dag\neq S^+_{kl}$ and 
$(S^-_{\bar k l})^\dag\neq S^+_{
\bar kl}$ if $\sigma_k\neq \sigma_l$. Thus,    $2S^\mu_{kl}$ and $2S^\mu_{\bar kl}$  have still integer eigenvalues $2m$ with  $|m|=0,\frac{1}{2},1$, for $\mu=x,y,z$ (here $S^x=\frac{S^++S^-}{2}$, $S^y=\frac{S^+-S^-}{2i}$) 
while   $S^{\pm}_{kl}$  and $S^{\pm}_{\bar kl}$ are  ladder-type operators, with   $S^+_{kl}S^-_{kl}$ and 
$S^+_{\bar kl}S^-_{\bar kl}$  having {\it integer} eigenvalues $S(S+1)-m(m-1)$ with $S=0,\frac{1}{2},1$,  $|m|\leq S$, i.e.\ $0,1,2$. 
The pair condensates \eqref{State}  correspond to $S=0$ $\forall\,k,l$,  
due to Eq.\ \eqref{19}.

Besides, the $\frac{3}{2}n$ conserved ``diagonal'' operators 
\begin{equation}
\begin{split}
S_{k}^+&=\tfrac{Q_{\bar k \bar k}}{2\sigma_k}=a^\dag_k a_{\bar k},\;\;
S_{k}^-=\tfrac{Q_{kk}}{2\sigma_k}=a^\dag_{\bar k} a_k,\\
S_{k}^z&=\tfrac{Q_{\bar k k}}{2\sigma_k}=\tfrac{1}{2}(a^\dag_k a_k-a^\dag_{\bar k}a_{\bar k})\,,
\label{B11}
\end{split}
\end{equation}
which do not depend on the $\sigma_k$,  also satisfy $SU(2)$ commutation relations. These operators are actually conserved in any paired state of the form \eqref{Pairing},  which include  in particular pair condensates, and lead to hermitian angular momentum-like operators $S^\mu_k$, $\mu=x,y,z$. 

In the bosonic case, defining first 
  \begin{equation}\tilde Q_{kl}=i \tfrac{Q_{kl}}{\sqrt{\sigma_k\sigma_l}}=i\left(\sqrt{\tfrac{\sigma_k}{\sigma_l}}b^\dag_k b_l-\sqrt{\tfrac{\sigma_l}{\sigma_k}}b^\dag_l b_k\right)\,,\label{B12}\end{equation} for $k<l$, with $Q_{kl}$ the  operators  \eqref{Qconservedb}, the triad 
  \begin{equation}(S^1,S^2,S^3)=(\tilde Q_{jk},\tilde Q_{kl},\tilde Q_{jl})\label{B13}\end{equation}   satisfies standard angular momentum commutation relations  $[S^\mu,S^\nu]=i\epsilon^{\mu\nu\sigma}S^\sigma$ $\forall\,j<k<l$ ($\epsilon^{\mu\nu\sigma}$ is the Levi-Civita symbol).  Though non-hermitian for $\sigma_k\neq \sigma_l$, all $\tilde Q_{kl}$ have   
{\it integer} eigenvalues $m\in \mathbbm{Z}$ $\forall\,k<l$, as they are connected to an angular momentum operator, or equivalently, a  two-mode  boson number difference \cite{RK.09}, by the  similarity transformation 
 \eqref{transQ}: 
$\tilde Q_{kl}=e^{-h}\tilde Q^0_{kl}e^h$, with \begin{equation}
    \tilde Q^0_{kl}=i(b^\dag_k b_l-b^\dag_l b_k)=x_l p_k-p_l x_k= b^\dag_{\tilde k}b_{\tilde k}-b^\dag_{\tilde l}b_{\tilde l}\,.
\end{equation} 
Here $x_j=\frac{b_j+b^\dag_j}{\sqrt{2}}$, $p_j=\frac{b-b^\dag}{\sqrt{2}i}$ are coordinate-momentum operators, such that $\tilde Q^0_{kl}$ is an angular momentum operator, and   $b_{\tilde k}=\frac{b_k+ib_l}{\sqrt{2}}$, $b_{\tilde l}=\frac{b_l+ib_k}{\sqrt{2}}$ standard boson annihilation operators.   Pair condensates lead to $0$ spin for all triads  as  $Q_{kl}|m\rangle_2=0$  $\forall\,k<l$ (Eq.\ \eqref{19}).  

\section{Arguments for the Conjecture\label{ApC}}
We will consider even $N$-particle states in a sp space of even finite dimension $n$, having a full rank one-body DM ${\bm \rho}^{(1)}$, i.e.\ $\bm\rho^{(1)}>0$, such that there are  no empty levels. For fermions we will also assume  no fully occupied levels, i.e.\,  $\bm \rho^{(1)}(\mathbbm 1-\bm \rho^{(1)})>0$, implying $2\leq N\leq n-2$.  

Any two-particle state  can be written as  $A^\dag|0\rangle$, with $A^\dag$ a pair creation operator \eqref{A}--\eqref{Schmidt}, having full rank ($\sigma_k>0\,\forall\,k$) if complying with previous conditions. The  number of linearly independent conserved one-body operators  for any such state is $L_n$, according to Eq.\ \eqref{L}, 
comprising $\hat N$ and the  $\frac{n(n\pm 1)}{2}$  operators  $Q_{ij}$, Eq.\ \eqref{Qconserved}. Since all $Q_{ij}$   fulfill $Q_{ij}|0\rangle=0$ and commute with $A^\dag$ (Eq.\ \eqref{CQA}),  they are also conserved for any $m$-pair condensate $(A^\dag)^m|0\rangle$.
We provide here arguments supporting that no other $2m$-particle state satisfying previous conditions has a larger number of conserved one-body operators. 

A $2m$-particle state can be written as $|\psi\rangle=\Gamma_m^\dag|0\rangle$, with $\Gamma_m^\dag$ a $2m$-particle creation operator. As any one-body operator $Q$ satisfies $Q|0\rangle=0$ and $[Q,\Gamma^\dag_m]=\Gamma_{m_Q}^\dag$, with $\Gamma_{m_Q}^\dag$ also a $2m$-particle creation operator, it follows that $[Q,\Gamma_m^\dag]=\lambda \Gamma_m^\dag$ if conserved. 
Then, by the same arguments given in App. B, the associated operator $\tilde Q=Q-\frac{\lambda}{2m}\hat N$, satisfying $\tilde Q|\psi\rangle=0$, must fulfill $[\tilde Q,\Gamma^\dag_m]=0$. For $m\geq 2$ states complying with previous conditions, this implies a number of constraints on $\tilde Q$ which is normally larger than those for a pair creation operator, entailing fewer conserved operators   
unless $\Gamma_m^\dag$ is a function of $A^\dag$, i.e.\ $\Gamma_m^\dag\propto (A^\dag)^m$ for $2m$-particle states.  

In fact, in contrast with two-particle states, typical random $2m$-particle states with $m\geq 2$ (and $m\leq n/2-2$ for fermions) have just one conserved one-body operator, i.e.\ the particle number $\hat N$, as verified numerically. 
The peculiarity of the $m$-pair condensates \eqref{State} is that they have the same number of conserved one-body operators as two-particle states, for {\it  any} $m$. 
 Special $2m$-particle states may have, of course, other  conserved one-body operators in addition to $\hat N$, but their number is seen to be lower than $L_n$ in typical families. For example, as shown in section \ref{IIC}, paired states of the general form \eqref{Pairing} have just $L_n^p=3n/2+1$ ($n/2+1$) conserved one-body operators for fermions (bosons) if $m\geq 2$ (and $m\leq n/2-2$ for fermions),  which is lower than $L_n$.  

If we now consider a ``product''of pair condensates \begin{equation}|\psi\rangle\propto (A^\dag)^{m_A} (B^\dag)^{m_B}|0\rangle\end{equation} with $A^\dag$ and $B^\dag$ acting on orthogonal sp subspaces ${\cal S}_A$, ${\cal S}_B$ of finite even dimensions $n_A$ and $n_B=n-n_A$, 
just those for each  condensate will  be conserved, since any one-body operator $Q_{AB}=\sum_{i\in {\cal S}_A,j\in{\cal S}_B}Q^{AB}_{ij}c^+_i c_j$ destroying one particle in $B$  and creating one in $A$ (or viceversa) will have a non-zero covariance   if $\bm\rho^{(1)}>0$ (and also $\mathbbm{1}-\bm\rho^{(1)}>0$ for fermions): 
In such state $\rho^{(1)}_{i_A,j_B}=0$ and hence 
$\langle Q_{AB}\rangle=0$, whereas, using the natural orbitals 
$\rho^{(1)}_{i_A,i'_A}=\delta_{ii'}f^A_i$, $\rho^{(1)}_{j_B,j'_B}=\delta_{jj'}f^B_{j}$, 
we obtain $\langle Q_{AB}^\dag Q_{AB}\rangle=\sum_{i,j}
|Q^{AB}_{ij}|^2 (1\mp f^A_i)f^j_B>0$ for fermions (bosons). 
Hence the number of one-body conserved operators will be $L_{n_A}+L_{n_B}=L_n-(n_A n_B-1)<L_n$ for both fermions and bosons if $n_A, n_B\geq 2$.   

As a third example, let us consider the $N=\frac{n}{2}$ state   
\begin{equation}
    |\psi\rangle=(\alpha c^\dag_1\ldots c^\dag_{\frac{n}{2}}+
\beta c^\dag_{\frac{n}{2}+1}\ldots c_{n})|0\rangle\,,\label{GHZ}\end{equation} 
with $\alpha\beta>0$ and $n\geq 8$, which also leads to a full rank $\bm \rho^{(1)}$ (with eigenvalues $|\alpha|^2$ and $|\beta|^2$, $\frac{n}{2}$-fold degenerate). This state is   a superposition of two SDs or permanents in orthogonal sp spaces,  and can be considered as a fermionic or bosonic analogue of a generalized GHZ (Greenberger-Horne-Zeilinger)-type state 
$\alpha |00\ldots\rangle+\beta|11\ldots\rangle$ \cite{DVC.00,HHHH.09}. The number of linearly independent conserved one-body operators for fermions (F) and bosons (B) is  
\begin{equation}
L^{g}_n=
\left\{^{n^2/2-1\;\;\;(F)}_{\;\;n-1\;\;\;\;\;\;(B)}\right. \;  <L_n\,,
\label{ghzl}
\end{equation}
i.e.\ $\hat N$ and the $n-2$ operators 
$Q_{i}=n_i-n_1$,  $Q_{i+\frac{n}{2}}=n_{i+\frac{n}{2}}-n_{\frac{n}{2}}$ for $n_i=c^\dag_i c_i$ and $i=2,\ldots,\frac{n}{2}$, with fermions having in addition the $n(\frac{n}{2}-1)$ operators $Q_{ij}=c^\dag_j c_i$ for $i<j$ belonging to the same half. They all satisfy $Q_{\alpha}|\psi\rangle=0$. This leads to $L_n-L_n^g\geq 2 + \frac{n}{2}$ ($n\geq 8$).    

The states \eqref{ghzl} are particular cases of 
\begin{equation}
|\Psi\rangle \propto \sum_{m_{1}\cdots m_{d}}\Gamma_{m_{1}\cdots m_{d}} (A_{1}^{\dag})^{m_1} \cdots (A_{d}^{\dag})^{m_d}|0\rangle\,,
\label{group}
\end{equation}
where $A_p^\dag = \prod_{i=1}^{n_p} (a_{pi}^\dagger)^{l_{pi}}$, with $\sum_{p=1}^d n_p = n$, $l_{pi}=1$ ($\geq 1$), $m_p=0,1$ ($\geq 0$) for fermions (bosons), and a fixed total particle number is assumed. Then,  there are  
\begin{equation}
L' = \sum_{p=1}^d (n_p^2-1) + 1  <L_n\,,
\end{equation}
conserved one-body operators for fermions: the particle number $\hat N$ and the special operators 
\begin{equation}
Q^p_{ij} = (a_{pi}^\dag a_{pi} - \tfrac{\hat{N}_p}{n_p}) \delta_{ij} + a_{pi}^\dag a_{pj} (1-\delta_{ij})\,,
\label{Qconserved3}
\end{equation}
with $\hat{N}_p = \sum_i a_{pi}^\dag a_{pi}$ ($\sum_i Q^p_{ii} = 0$), for $i,j=1,\ldots,n_p$.  
For $d=n/2$ and $n_p=2$, we recover the paired states \eqref{Pairing},  with $L' = 3n/2+1=L_n^p$ as expected, while for $d=2$ and $n_p=n/2$, we recover the previous GHZ-like states, where $L'=n^2/2-1=L^g_n$. 

For fixed $d$, the maximum value of $L'$ is reached for $n_p = n/d \ \forall \ p$, in which case $L'= n^2/d-d+1$. For $d=1$ we obtain the fully occupied SD and $L'= n^2$ as expected, i.e., all one body operators are conserved. For $d \geq 2$, this $L'$ is maximum for $d=2$, which corresponds to the GHZ-like states, Eq.\  \eqref{ghzl}, such that $L'$ never exceeds $L_n$ for states complying with $\bm\rho^{(1)}(\mathbbm{1}-\bm{\rho}^{(1)})>0$. 

In the bosonic case, the special conserved operators \eqref{Qconserved3} become instead  
\begin{equation}
Q^p_{i} = a_{pi}^\dag a_{pi} - \tfrac{l_{pi}}{n_p} \sum_k \tfrac{a_{pk}^\dag a_{pk}}{l_{pk}} \,.
\label{Qconserved4}
\end{equation}
 Those with $i \neq j$ are no longer conserved, and then 
 \begin{equation}
L' = \sum_{p=1}^d (n_p-1) + 1 = n-d+1 < L_n\,.
\end{equation}
Remarkably, for $d=1$, i.e., a permanent boson state (with $\bm\rho^{(1)}>0$),  $L'=n$ (we can just take $Q_i=c^\dag_i c_i$ for $i=1,\ldots,n$) and then it also has a smaller number of conserved one-body operators than the pair condensate.   
 
\section{proof of Proposition 2\label{ApD}}

First, we notice that the eigenvalues of the covariance matrix  \eqref{C20}, i.e.\ of the two-body DM $\bm\rho^{(2)}$,  are analytical for the plain state $|m_0\rangle_2$, being all nonzero  for $m \geq 2$  in both fermion and boson systems \cite{CR.24}, implying that $|m_0\rangle_2$ has no strictly conserved quantities linear in $c_i c_j$. This entails that there are neither conserved quantities of this form in all states \eqref{State} for $m\geq 2$, due to Eq.\ \eqref{transQ}.  

Regarding the operators linear in $c_i^\dag c_j^\dag$, in the fermionic case,  they cannot be conserved for $m\leq n/2-2$, since  owing to Eq. \eqref{Statebis}, the matrix \eqref{C02} becomes equivalent to \eqref{C20} in the corresponding hole condensate, hence lacking any zero eigenvalue. And in the bosonic case, the covariance \eqref{C02} reads 
\begin{eqnarray*}
\bar{\rho}_{ij,i'j'}^{(2)} &=& \delta_{ii'}\delta_{jj'} + \delta_{ij'}\delta_{ji'} \\
&+& \delta_{ii'}\rho_{jj'}^{(1)}+\delta_{ji'}\rho_{ij'}^{(1)}+\delta_{ij'}\rho_{ji'}^{(1)}+\delta_{jj'}\rho_{ii'}^{(1)}+\rho_{ij,i'j'l}^{(2)}\,.
\end{eqnarray*}
Then it is  always positive definite and hence need not be considered for seeking conserved operators.

\section{proof of Theorem 2\label{ApE}}

We  consider $m\geq 2$ (and $m\leq \frac{n}{2}-2$ for fermions).  Then, using commutation properties it can be proved that for a two-body Hamiltonian conserving the particle number, 
\begin{eqnarray}
\!\!\!\! H |m\rangle\!&=&\! m (A^\dag)^{m-2} \left( \tfrac{m-1}{2} [[H,A^\dag],A^\dag]\!+\! A^\dag H A^\dag \right)|0\rangle,\;\;\;\;\;\;
\label{Hcon}
\end{eqnarray}
implying that Eq. \eqref{eigen} is fulfilled iff (see below)
\begin{equation}
(\tfrac{m-1}{2} [[H,A^\dag],A^\dag] + A^\dag H A^\dag) |0\rangle = \alpha_m (A^\dag)^2|0\rangle,
\label{Hcon2}
\end{equation}
where $\alpha_m = \lambda_m/m$. We can always write
\begin{equation}
H A^\dag |0\rangle = (\alpha_1 A^\dag - \gamma A^\dag_{\perp}) |0\rangle\,,
\label{extracon}
\end{equation}
with $\langle 0| A_\perp A^\dag |0\rangle = 0$ and then, Eq. \eqref{Hcon2} becomes
\begin{equation}
(\tfrac{m-1}{2} [[H,A^\dag],A^\dag] \!-\! \gamma A^\dag A^\dag_\perp) |0\rangle \!=\! (\alpha_m-\alpha_1) (A^\dag)^2 |0\rangle.
\label{Hcon3}
\end{equation}
It is convenient now to define
\begin{equation}
\tilde{H} = H - \tfrac{\alpha_m-\alpha_1}{4(m-1)} \hat{N}^2,
\label{tildeH}
\end{equation}
implying
\begin{equation}
[[\tilde{H},A^\dag],A^\dag] |0\rangle = \gamma A^\dag A^\dag_\perp |0\rangle.
\label{Hcon33}
\end{equation}
We will first solve the homogeneous equation ($\gamma = 0$) and then we will find a particular solution for $\gamma \neq 0$.

Since the set of $O_{ij} = (\bm{c}^{\dagger}{\bf A}^t)_i c_j$ form a basis of one-body operators ($c_i^\dag = \sum_j A^{-1}_{ij} (\bm{c}^{\dagger}{\bf A}^t)_j$), it is convenient to write the homogeneous solution $\tilde{H}_h$ as follows,
\begin{eqnarray}
\tilde{H}_h &=& \tilde{h} +  \sum_{i,j,k,l} U_{ij,kl} O_{ij} O_{kl} \\ &=&\tilde{h} + \tfrac{1}{4} \sum_{i,j,k,l} \sum_{\sigma \sigma' = \pm} U^{\sigma \sigma'}_{ij,kl} Q^{\sigma}_{ij} Q_{kl}^{\sigma'}\,,
\label{tildeH22}
\end{eqnarray}
with $\tilde{h}$ a one body operator and $Q^{\pm}_{ij} = O_{ij} \pm O_{ji} = \pm Q^{\pm}_{ji}$.

Taking into account that $[Q_{ij}^{\pm},A^\dag] = 0$, we can see that Eq. \eqref{Hcon33} only imposes restrictions for $U^{\mp \mp}_{ij,kl} = \mp U^{\mp \mp}_{ji,kl} = \mp U^{\mp \mp}_{ij,lk} = U^{\mp \mp}_{kl,ij}$ respectively, and it leads to
\begin{equation}
\sum_{i,j,k,l} U^{\mp \mp}_{ij,kl} (\bm{c}^{\dagger}{\bf A}^t)_i (\bm{c}^{\dagger}{\bf A}^t)_j (\bm{c}^{\dagger}{\bf A}^t)_k (\bm{c}^{\dagger}{\bf A}^t)_l = 0,
\label{condU}
\end{equation}
implying
\begin{equation}
U^{\mp \mp}_{ij,kl} = \pm (U^{\mp \mp}_{ik,jl} + U^{\mp \mp}_{il,kj}), 
\label{condUS}
\end{equation}
where the upper sign corresponds to fermions and the lower one to bosons as always.

Thus, we have
\begin{eqnarray*}
\hat{U}^{\mp \mp} &\!:=\!& \tfrac{1}{4} \sum_{i,j,k,l} U_{ij,kl}^{\mp \mp}  Q_{ij}^{\mp} Q_{kl}^{\mp} = \sum_{i,j,k,l} U_{ij,kl}^{\mp \mp}  O_{ij} O_{kl}  \\ 
&\!=\!& \tfrac{1}{3} \sum_{i,j,k,l} U_{ij,kl}^{\mp \mp}  O_{ij} O_{kl} \!+\! U_{ik,jl,}^{\mp \mp}  O_{ik} O_{jl} \!+\! U_{il,kj}^{\mp \mp}  O_{il} O_{kj} \;\;\;\;\ \\
&\!=\!& \tfrac{1}{3} \sum_{i,j,k,l} U_{ik,jl}^{\mp\mp}(O_{ik}O_{jl}\pm O_{ij}O_{kl}) \\
&\!+\!& \tfrac{1}{3} \sum_{i,j,k,l} U_{il,kj}^{\mp\mp}(O_{il}O_{kj}\pm O_{ij}O_{kl}).
\end{eqnarray*}
Using commutation relations, it can be easily shown that $O_{ik}O_{jl}\pm O_{ij}O_{kl} =  h_1 + (\bm{c}^{\dagger}{\bf A}^t)_i c_lQ_{jk}^{\pm}$ whereas $O_{il}O_{kj}\pm O_{ij}O_{kl} = h_2$ with $h_1$ and $h_2$ one body terms, and hence we finally obtain that $\tilde{H}_h$ has the form
\begin{equation}
\tilde{H}_h = \tilde{h}^{\prime} +  \sum_{i,j,k,l} \tilde{U}_{ij,kl} c^\dag_i c_j Q_{kl}.
\label{tildeH23}
\end{equation}
with $\tilde{h}^{\prime}$ a one body term, for both fermions and bosons. 

Regarding the particular solution, we can take
$\tilde{H}_p = \gamma A^\dag B$ with $B^\dag$ a two particle creation operator satisfying  $[[B,A^\dag],A^\dag] = A^\dag_\perp$ (there is always a choice of $B$ such that this is fulfilled). Thus,   $\tilde{H}$ has the form
\begin{equation}
\tilde{H} = \tilde{h}^{\prime} + \gamma A^\dagger B +  \sum_{i,j,k,l} \tilde{U}_{ij,kl} c^\dag_i c_j Q_{kl}.
\label{tildeH2}
\end{equation}
The one body term is obtained by replacing the original Hamiltonian $H$ in \eqref{extracon} leading to
\begin{equation}
\begin{split}
\!\!\!\!H&=\alpha \hat{N} + \beta \hat{N}^2 + \gamma [(1+m) A^\dag B + (1-m) B A^\dag] \\
&\;\;\;\;+  \sum_{i,j} h_{ij} Q_{ij} + \sum_{i,j,k,l} \tilde{U}_{ij,kl} c^\dag_i c_j Q_{kl}.
\label{tildeH2a}
\end{split}
\end{equation}
Finally, it can be easily shown that
\begin{equation*}
(1+m) A^\dag B + (1-m) B A^\dag = 1+m -\tfrac{1}{2} \sum_{i,j} (Q_{ij}^{B})^\dag Q_{ij}\,,
\end{equation*}
with $Q^B_{ij} =(\bm{c}^{\dagger}{\bf B}^t)_i c_j \pm(\bm{c}^{\dagger}{\bf B}^t)_{j} c_i$ the conserved quantities associated to the state $B^\dag |0\rangle$, implying that $H$ has the final form
\begin{equation}
H = \alpha \hat{N} \!\!+\!\! \beta \hat{N}^2 \!\!+\!\! \sum_{i,j} h_{ij} Q_{ij} \!\!+\!\! \sum_{i,j,k,l} V_{i,j,k,l} c^\dag_i c_j Q_{kl}.
\label{tildeH2aa}
\end{equation}

The last step of the proof is to demonstrate that Eq. \eqref{eigen} implies \eqref{Hcon2}. In the bosonic case this is obvious since the creation operators do not have null space. In the fermionic case, for $m=2$  this is also obvious and then we will consider, for instance, $m=3$. In this case, Eq. \eqref{eigen} has the form
\begin{equation}
A^\dag C^{(4) \dag} |0\rangle = 0,
\label{eigen2}
\end{equation}
where
\begin{equation}
C^{(4) \dag} = \tfrac{m-1}{2} [[H,A^\dag],A^\dag] + A^\dag H A^\dag - \alpha_m (A^\dag)^2.
\end{equation}
is a four particle creation operator. Applying $\bar{A}$ to both members of Eq. \eqref{eigen2} and using \eqref{comAbar} we arrive at
\begin{equation}
(1-\tfrac{8}{n}) C^{(4) \dag} + A^\dag \bar{A} C^{(4) \dag} |0\rangle = 0.
\label{eigen3}
\end{equation}
Thus, since $m\leq \frac{n}{2} -2$, i.e. $\frac{n}{2} \geq 5$ in this case (impliying $1-\tfrac{4}{n/2} \neq 0$), Eq. \eqref{eigen3} implies that $C^{(4) \dag} = A^\dag B^\dag$ with $B^\dag |0\rangle \propto \bar{A} C^{(4) \dag} |0\rangle$ a two particle creation operator. Replacing in \eqref{eigen2} we have $A^{\dag2} B^\dag |0\rangle = 0$ and then $B^\dag = 0$ due to Proposition 2. This implies $C^{(4)} = 0$ and then Eq. \eqref{Hcon2}. The proof is similar for $4 \leq m \leq \frac{n}{2} - 2$.\qed


%
\end{document}